\documentclass[prb,twocolumn,superscriptaddress,showpacs,amsmath,amssymb]{revtex4}
\usepackage{float}
\usepackage{amsmath}
\usepackage{bm}
\usepackage[usenames]{color}
\usepackage{verbatim}
\usepackage{graphicx}
\newcommand{\be}{\begin{equation}}
\newcommand{\ee}{\end{equation}}   
\newcommand{\bea}{\begin{eqnarray}}
\newcommand{\eea}{\end{eqnarray}}
\newcommand{\phrl}[1]{Phys.~Rev.~Lett. {\bf #1}}
\newcommand{\phrb}[1]{Phys.~Rev.~B {\bf #1}}
\newcommand{\phrx}[1]{Phys.~Rev.~X {\bf #1}}
\newcommand{\cmat}[1]{arXiv:{\bf #1}}

\newcommand{\jpcm}[1]{J.~Phys.:Condens.~Matter.{\bf #1}}
\newcommand{\bib}{\bibitem}
\newcommand{\lb}{\left[}
\newcommand{\rb}{\right]}
\newcommand{\lp}{\left(}
\newcommand{\rp}{\right)}
\newcommand{\lf}{\left\{}
\newcommand{\rf}{\right\}}

\renewcommand{\k}{\mathbf{k}}

\newcommand{\q}{\mathbf{q}}
\newcommand{\tk}{\tilde{k}_{s'}}
\newcommand{\tkz}{\tilde{k}_{z,s'}}
\newcommand{\tOmega}{\widetilde{\Omega}}
\newcommand{\tQ}{\widetilde{Q}}

\begin{document}
\title{Doping and tilting on optics in noncentrosymmetric multi-Weyl semimetals}

\author{S. P. Mukherjee}
\affiliation{Department of Physics and Astronomy, McMaster University, Hamiltion, Ontario, Canada L8S 4M1}

\author{J. P. Carbotte}
\affiliation{Department of Physics and Astronomy, McMaster University, Hamiltion, Ontario, Canada L8S 4M1}
\affiliation{Canadian Institute for Advanced Research, Toronto, Ontario, Canada M5G 1Z8}

\begin{abstract}
Weyl semimetal (WSM) feature tilted Dirac cones and can be type I or II depending on the magnitude of the tilt parameter ($C$). The boundary between the two types is at $C=1$ where the cones are tipped 
and there is a Lifshitz transition. The topological charge of a WSM is one. In multi-Weyl it can be two or more depending on the value of the winding number $J$. We calculate the absorptive part of the 
AC optical conductivity both along the tilt direction ($\sigma_{zz}$) and perpendicular to it ($\sigma_{xx}$) as a function of the tilt ($C$) and chemical potential ($\mu$). For zero tilt there is a 
discontinuous rise in both $\sigma_{xx}$ and $\sigma_{zz}$ at photon energy $\Omega=2\mu$ followed by the usual linear in $\Omega$ law for $\sigma_{xx}$ at $J=1,2$ and $\sigma_{zz}$ at $J=1$. For $J=2$ 
and $\sigma_{zz}$ the interband background is constant rather than linear in $\Omega$. For type I there is a readjustment of optical spectral weight as the tilt is increased. The absorption starts from
zero at $2\mu/(1+C)$ and then rises in a quasilinear fashion till it merge with the usual undoped untilted interband background at $2\mu/(1-C)$. The discontinuous rise at twice the chemical potential of 
the untilted case is lost. For type II the interband background of the undoped untilted case is never recovered. For noncentrosymmetric materials the energies of a pair of opposite chirality Weyl nodes 
become shifted by $\pm Q_{0}$ and this leads to two separate absorption edges corresponding to the effective chemical potential of each of the two nodes at $2(\mu+\chi Q_{0})$ depending on chirality 
$\chi=\pm$. We provide analytic expressions for the conductivity in this case which depend only on the ratio $Q_{0}/\mu$ and tilt when plotted against $\Omega/\mu$. The signature of finite energy shift 
$Q_{0}$ is more pronounced for $\sigma_{zz}$ and $J=2$ than for the other cases.
\end{abstract}


\maketitle

\section{Introduction}
\label{sec:I}

Weyl fermions are known to exist in many different classes of semimetals. Initially suggested in the pyrochlore iridates, Rn$_2$In$_2$O$_7$\cite{Wan}, this was followed with theoretical investigations 
in the noncentrosymmetric transition metal monophosphides \cite{Dai} which were soon verified in experiments for TaAs \cite{Lv,Xu,Ding,Yang} and other related materials\cite{Shi,Alidoust} including the 
time-reversal(TR) symmetry breaking compound YbMnBi$_2$\cite{Borisenko}. Weyl nodes come in pairs of opposite chirality and result when the degeneracy of a doubly degenerate Dirac point is lifted through 
broken inversion or TR symmetry. The Weyl cones can be tilted with respect to the energy axis and type I or II Weyl nodes result depending on the magnitude of the tilt ($C$). For $C$ less than one in units 
of the appropriate Fermi velocity we have type I and for $C>1$ type II \cite{Soluyanov} (overtilted). For type I in the undoped material the density of state at the Fermi surface remains zero but for 
type II it becomes finite as electron and hole pockets have formed \cite{Huang,Bruno,Singh,Deng,Liang,Jiang,Dil,Kaminski,Khim,Kopernik,Haubold,Autes,Belopolski}. Hybrid Weyl semimetals\cite{Zhang} have 
also been considered where the tilt of the opposite chirality node can be different and indeed one type I with the other type II. The topological charge of a Weyl node can be greater than one for winding 
numbers $J$ two and above and these semimetals are referred to as multi-Weyl\cite{Fang,Bernevig,Yao,Shen,Sun,Lu}. The absorptive part of the longitudinal AC optical conductivity 
$\Re\sigma_{ii}(\Omega)$ with $i=x,y$ as a function of photon energy provides direct information on the dynamics of the Dirac and Weyl fermions\cite{Nicol,Tabert,Carbotte,Mukherjee,Mukherjee1,Pesin} as 
experiments have confirmed\cite{Chen,Sushkov,Xiao,Neubauer,Timusk,Chinotti}. The optical conductivity in multi-Weyl semimetal (mWSM) has been considered by Ahn, Mele and Min \cite{Mele}. In this paper we 
considered the AC optical conductivity along the direction of the tilt $\Re\sigma_{zz}(\Omega)$  and perpendicular to it $\Re\sigma_{xx}(\Omega)$ with particular emphasis on the effect of tilt and of 
doping away from charge neutrality. We also include in our continuum Hamiltonian, terms that deal with both TR and inversion symmetry breaking\cite{Shan,Zyuzin}. The transport properties of mWSM  
including the effect of anisotropic residual scattering, short range and charged impurities within a Boltzmann approximation have been considered by Park et. al.\cite{Woo} and the magnetoconductivity by 
Sun and Wang\cite{Wang} following closely previous work on ordinary Weyl\cite{Ashby}.

The paper is structured as follows. In section II we specify the model continuum Hamiltonian on which all our work is based. It includes a pair of multi-Weyl node of opposite chirality and terms which 
break time reversal and inversion symmetries. The first displaces the Weyl nodes in momentum space by $\pm\bf{Q}$ and the second displaces them in energy by a shift $\pm Q_{0}$. The Green's function is 
specified and the current $J_{x}$ (perpendicular to the tilt direction) and $J_{z}$ (parallel to the tilt direction) are computed from the Hamiltonian. From this information the absorptive part of the 
AC optical conductivity $\Re\sigma_{xx}(\Omega)$ and $\Re\sigma_{zz}(\Omega)$ as a function of photon energy are calculated from a Kubo formula. In section III we reduce the expression for both 
$\Re\sigma_{xx}(\Omega)$ and $\Re\sigma_{zz}(\Omega)$ to simple analytic formulas which depend on the tilt and on the doping through the value of the chemical potential and on the energy shift $Q_{0}$ 
of the Weyl nodes. The displacement in momentum $\bf{Q}$ of the two nodes, which is known to play a key role for the real part of the DC anomalous Hall conductivity, drops out of the expressions for the 
absorptive part of the conductivity. In section IV we present results. We start with the centrosymmetric case and show results for $\Re\sigma_{zz}(\Omega)$ as a function of variable tilt comparing the 
case $J=1$ (Weyl) with $J=2$ (multi-Weyl) and highlight the difference between type I and type II Weyl. In section V this is followed with a series of results for noncentrosymmetric materials for which 
$Q_{0}$ is non-zero. Finite $Q_{0}$ leads to two steps in both $\Re\sigma_{xx}(\Omega)$ and $\Re\sigma_{zz}(\Omega)$ which become modified as the tilt is increased. A comparison of $\Re\sigma_{zz}(\Omega)$ 
with $\Re\sigma_{xx}(\Omega)$ is presented for a fixed illustrative value of $Q_{0}=0.5$ for the case of $J=1$ and both type I and II are considered. Similarities and differences are emphasized, as is 
the low photon energy region. Next a comparison of results for the $\Re\sigma_{xx}(\Omega)$ and $\Re\sigma_{zz}(\Omega)$ in the case of $J=1$ with emphasis on the low photon energy regime is presented. 
Only type I case is considered and the value of $Q_{0}$ is varied with a view to understand how the conductivity reflects this parameter. Of particular interest is the appearance of a particular region 
of photon energies in which only the negative chirality node is contributing. This is followed by a comparison of the $J=1$ to the $J=2$ case both for $\Re{\sigma_{zz}}(\Omega)$ for $C$ fixed at 0.5 
(type I) and another at 1.5 (type II) while the energy shift between the two Weyl nodes of opposite chirality is varied. In section VI we provide a summary and conclusion.

\section{Formalism}
\label{sec:II}

\begin{figure}[H]
\centering
\includegraphics[width=1.5in,height=1.5in, angle=0]{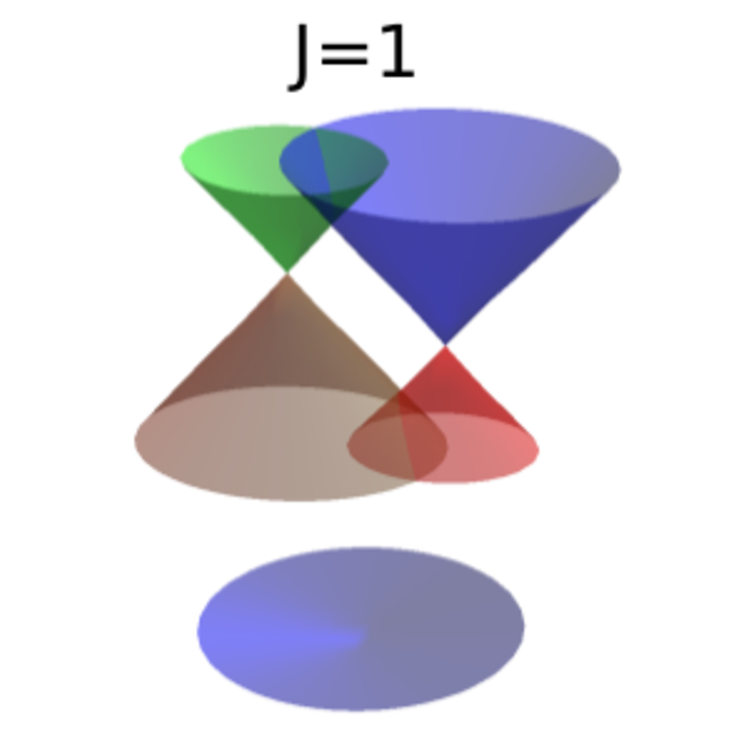}
\includegraphics[width=1.5in,height=1.5in, angle=0]{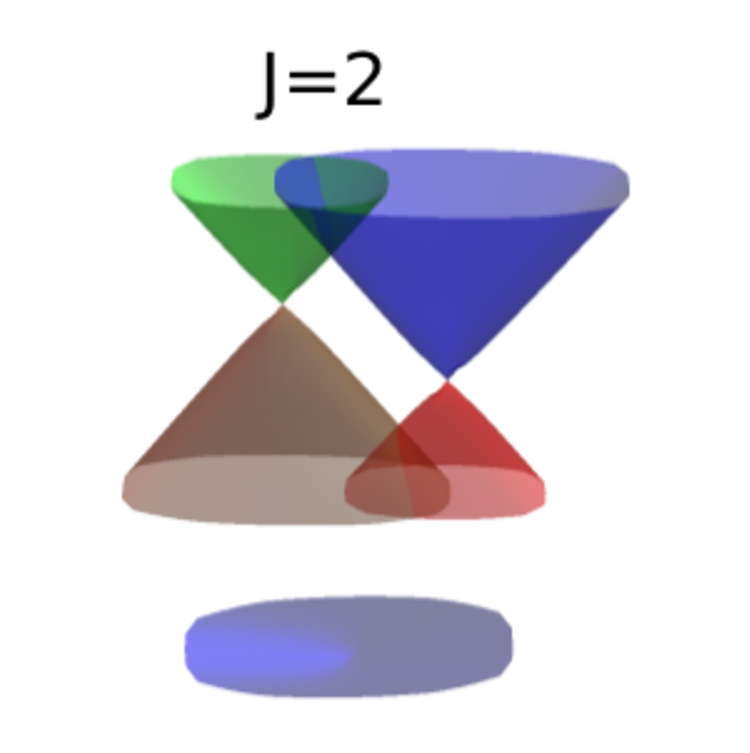}
\caption{(Color online) Effect of both broken time-reversal symmetry as well as inversion symmetry on untilted mWSM with different winding number $J$. The one on the left is for $J=1$, and on the right 
for $J=2$. We see that the cross section gets elongated from $J=1$ to $J=2$ as shown.} 
\label{Fig1}
\end{figure}

We begin with the minimal continuum Hamiltion for an isolated Weyl node of chirality $s'$ with both broken TR and inversion symmetry. Additionally we assume that the winding number associated with the 
Weyl node is $J$. The broken time reversal symmetry displaces the Weyl cone in momentum space by an amount $\pm \bf{Q}$ while the broken inversion symmetry shifts their energy by $\pm Q_{0}$\cite{Nicol,Shan,Zyuzin}.
\bea
&& \hat{H}_{s'}(\k)= C_{s'}(k_{z}-s' Q)+s' \biggl\{ v_{\perp} k_{0}\lp \tilde{k}^{J}_{-} \sigma_{+} + \tilde{k}^{J}_{+} \sigma_{-}\rp + \nonumber \\
&& v_{z}(k_{z}-s' Q)\sigma_{z}\biggr\}-s'Q_{0} 
\label{Hamiltonian-mWSM}
\eea
Here $s'=\pm1$ for Weyl nodes of opposite chirality. $C_{s'}$ describe the amount of tilting of the particular chiral node. The velocity $v_{\perp}$ is the effective velocity of the quasiparticles in the 
plane perpendicular to the $z$-axis while $v_{z}$ is the velocity along it and tilt $C_{s'}$ is normalized by this parameter. Here $k_{0}$ is a system dependent parameter having the dimension of momentum.
Also $\tilde{k}_{\pm}=k_{\pm}/k_{0}$ with $k_{\pm}=k_{x}\pm\imath k_{y}$ and $\sigma_{\pm}=\frac{1}{2}\lp\sigma_{x}\pm\imath\sigma_{y}\rp$. The Pauli matrices $\sigma_{i}$ where $i=x, y, z$ are defined 
as usually by,
\be
\sigma_{x}=\lp\begin{array}{cc}0 & 1\\ 1 & 0 \end{array} \rp, \sigma_{y}=\lp\begin{array}{cc}0 & -\imath\\ \imath & 0 \end{array} \rp, \sigma_{z}=\lp\begin{array}{cc}1 & 0\\ 0 & -1 \end{array} \rp.
\ee
The electronic energy dispersions corresponding to the above Hamiltonian are,
\bea
&& \epsilon_{s,s'}=C_{s'} \lp k_{z}-s'Q \rp-s'Q_{0}+ sv_{\perp} \times \nonumber \\
&& \sqrt{k^2_{0} \lp\frac{k^2_{x}+k^2_{y}}{k^2_{0}}\rp^J+ v^2_{0}\lp k_{z}-s'Q\rp^{2}}
\label{Dispersion}
\eea
where $s=\pm$ stands for conduction($+$) and valence($-$) bands and $v_{0}=v_{z}/v_{\perp}$. For a set of values of the parameters ($v_{\perp}=1,k_{0}=1, v_{0}=0.5,Q=2.5, Q_{0}=0.5$) we plot in 
Fig.[\ref{Fig1}] the energy dispersion for different winding number $J$. We see that now the conical cross section deviates from circular and evolves to elongated elliptical at higher $J$. The matrix 
Green's function corresponding to the above Hamiltonian is given by,
\be
\hat{G}_{s'}(k,z)= \lb I_{2}z-\hat{H}_{s'}(\k)\rb^{-1},
\label{GF-definition}
\ee
where $I_{2}$ is a $2\times 2$ unit matrix. It is straight forward to show that the Green's function can be written in the following form,
\bea
&& \hat{G}_{s'}(k,z)= \frac{1}{2}\sum_{s=\pm} \frac{1}{z-C_{s'}\tkz+s\tk+s'Q_{0}} \nonumber \\
&& \lp \begin{array}{cc}1-ss'(v_{z}\tkz/\tk) & -ss'v_{\perp}k_{0}(\tilde{k}^{J}_{-}/\tk) \\ -ss'v_{\perp}k_{0}(\tilde{k}^{J}_{+}/\tk) &  1+ss'(v_{z}\tkz/\tk)\end{array} \rp
\eea
where $\tkz=k_{z}-s'Q$ and $\tk=\sqrt{v^2_{\perp}k^2_{0} \tilde{k}^{J}_{+}\tilde{k}^{J}_{-}+v^2_{z}\tkz^2}$.
Now we outline the derivation of the conductivity $\sigma_{i,j}$ where $i,j=x, y, z$. For this we need the current-current correlation function which in matrix form is given by,
\bea
&& \hat{\Pi}_{ij}(\Omega_{m},\q)=T\sum_{\omega_{n}}\sum_{s'=\pm} \int \frac{d^3k}{(2\pi)^3} \nonumber \\
&&\hat{J}_{i,s'}\hat{G}_{s'}(\k+\q,\omega_{n}+\Omega_{m})\times \hat{J}_{j,s'}\hat{G}_{s'}(\k,\omega_{n}),
\eea
where $\Omega_{m}$ and $\omega_{n}$ are Matsubara frequencies, $\q$ is the internal momentum which we set to zero and $\hat{J}_{i,s'}$ is the current operator  which is defined as,
\be
\hat{J}_{i,s'}= e \frac{\partial\hat{H}_{s'}(\k) }{\partial k_{i}}.
\ee
Next we have to evaluate the sum over the internal Matsubara frequencies $\omega_{n}$ and after that we have to analytically continue the result by replacing $\imath\Omega_{m}$ with $\Omega+\imath \delta$, 
which will give us the retarded current-current correlation function $\Pi_{ij}(\Omega,0)$. The conductivity is then defined by the following formula,
\be
\sigma_{ij}(\Omega)=-\frac{\Pi_{ij}(\Omega,0)}{\imath\Omega}.
\ee
Following the above prescription we derive the expressions for the longitudinal conductivity $\sigma_{xx}(\Omega)$ and $\sigma_{zz}(\Omega)$ as shown below,
\bea
&& \hspace{-0.4cm}\sigma_{xx}(\Omega)=-\frac{\Pi_{xx}(\Omega,0)}{\imath\Omega} \nonumber\\
&& \hspace{-0.4cm}=\frac{\imath e^2 J^2 k^{-(2J-1)}_{0}v^2_{\perp}}{2\pi^2\Omega} \sum_{s'=\pm}\int^{\Lambda-s'Q}_{-\Lambda-s'Q} dk_{z} \int^{\infty}_{0} \frac{k^{2J-1}_{\perp}dk_{\perp}}{k} \times \nonumber\\
&& \hspace{-0.4cm}\biggl\{f(C_{s'}k_{z}+k_{0}k-\mu_{s'})-f(C_{s'}k_{z}-k_{0}k-\mu_{s'})\biggr\} \times \nonumber \\
&& \hspace{-0.4cm}(k^2_{0}k^2+v^2_{z}k^2_{z}) \times \lb \frac{1}{4k^2_{0}k^2-\Omega^2}+ \imath \pi \delta(4k^2_{0}k^2-\Omega^2)\rb ,
\label{sigma-xx}
\eea
\bea
&& \hspace{-0.4cm}\sigma_{zz}(\Omega)=- \frac{\Pi_{zz}(\Omega,0)}{\imath\Omega} \nonumber\\
&& \hspace{-0.4cm}=\frac{\imath e^2}{2\pi^2k_{0}\Omega} \sum_{s'=\pm}\int^{\Lambda-s'Q}_{-\Lambda-s'Q} dk_{z} \int^{\infty}_{0} \frac{k_{\perp} dk_{\perp}}{k} \times \nonumber \\
&& \hspace{-0.4cm}\biggl\{f(C_{s'}k_{z}+k_{0}k-\mu_{s'})-f(C_{s'}k_{z}-k_{0}k-\mu_{s'})\biggr\} \times \nonumber \\ 
&& \hspace{-0.4cm}\biggl\{(C^2_{s'}+v^2_{z})\lf k^2_{0}k^2-v^2_{z}k^2_{z}\rf-v^2_{\perp}(C^2_{s'}-v^2_{z})k^{2J}_{\perp}k^{-2(J-1)}_{0}\biggr\} \nonumber \\
&& \hspace{-0.4cm}\times \lb \frac{1}{4k^2_{0}k^2-\Omega^2}+ \imath\pi \delta(4k^2_{0}k^2-\Omega^2)\rb.
\label{sigma-zz}
\eea
Here $\Lambda$ is the cutoff, $k_{\perp}$ is the momentum perpendicular to $k_{z}$, $f(E)=(e^{E/T}+1)^{-1}$ is the Fermi function at finite temperature $T$ with $\mu$ the chemical potential and 
$\mu_{s'}=\mu+s'Q_{0}$ is the effective chemical potential for Weyl node with chirality $s'$. Also we note that $k_{\perp}$ and $k_{z}$ are related to each other by the equation,
\be
k=\sqrt{v^2_{\perp}(k_{\perp}/k_{0})^{2J}+v^2_{z}(k_{z}/k_{0})^2}.
\ee

\section{Derivation of longitudinal conductivities for finite $Q$ and $Q_{0}$}
\label{sec:III}

In this section we derive the expressions for the longitudinal conductivities $\sigma_{xx}(\Omega)$ and $\sigma_{zz}(\Omega)$ which are the central results of this article. Details of the derivation are 
presented in Appendix A and B so here we only state the results.

\subsection{Results for $\Re\sigma_{xx,s'}(\Omega)$}

I. For $0<C'_{s'}<1$ (type I tilting)
\bea
&& \frac{\Re\sigma_{xx,s'}(\Omega)}{\mu' e^2/8\pi}= 0,~~\text{where}~~ \tilde{\Omega}<\tOmega^{s'}_{L}\nonumber\\
&& =I^{1}_{s'}, ~~\text{where}~~ \tOmega^{s'}_{U}>\tilde{\Omega}>\tOmega^{s'}_{L}\nonumber\\
&& =I^{2}_{s'}, ~~\text{where}~~~~ \tilde{\Omega}>\tOmega^{s'}_{U},
\label{sigma-xx-I}
\eea

II. For $C'_{s'}>1$ (type II tilting)
\bea
&& \frac{\Re\sigma_{xx,s'}(\Omega)}{\mu' e^2/8\pi}= 0,~~\text{where}~~\tilde{\Omega}<\tOmega^{s'}_{L}\nonumber\\
&& =I^{1}_{s'},~~\text{where}~~\tOmega^{s'}_{U}>\tilde{\Omega}>\tOmega^{s'}_{L}\nonumber\\
&& =I^{3}_{s'},~~\text{where}~\tilde{\Omega}>\tOmega^{s'}_{U},
\label{sigma-xx-II}
\eea
where for any variable $a$, $a'=a/v_{z}$ and $\tilde{a}=a'/\mu'$. Also we have used the following shorthands,
\bea
&& I^{1}_{s'}=\frac{J}{24}(4+\frac{3}{C'_{s'}}+\frac{1}{C'^3_{s'}})\tilde{\Omega}-\frac{J}{4}(\frac{1}{C'_{s'}}+\frac{1}{C'^3_{s'}})|1+s'\tilde{Q_{0}}| \nonumber \\
&& +\frac{J}{2C'^3_{s'}\tilde{\Omega}}|1+s'\tilde{Q_{0}}|^2-\frac{J}{3C'^3_{s'}\tilde{\Omega}^2}|1+s'\tilde{Q_{0}}|^3 \nonumber\\
&& I^{2}_{s'}=\frac{J\tilde{\Omega}}{3}\nonumber\\
&& I^{3}_{s'}= \frac{J}{12}(\frac{3}{C'_{s'}}+\frac{1}{C'^3_{s'}})\tilde{\Omega}+ \frac{J}{C'^3_{s'}\tilde{\Omega}}|1+s'\tilde{Q_{0}}|^2 \nonumber\\
&& \tOmega^{s'}_{L}=2\biggl|\frac{1+s'\tQ_{0}}{1+C'_{s'}}\biggr|,\tOmega^{s'}_{U}=2\biggl|\frac{1+s'\tQ_{0}}{1-C'_{s'}}\biggr|.
\label{I's}
\eea
Note that the displacement in momentum of the two Weyl nodes has dropped out of the absorptive part of the conductivity in our clean limit calculations. In the inversion symmetric case , with $Q_{0}=0$ 
our expressions (\ref{sigma-xx-I}) to (\ref{I's}) reduce, as they must, to those of reference [\onlinecite{Mukherjee1}]. In this reference only the case for winding number $J=1$ was treated but we have 
verified here that for multi-Weyl with $J\ne1$, we need to only multiply by $J$ as was found in reference [\onlinecite{Mele}] for the no tilt case.

\subsection{Results for $\Re\sigma_{zz,s'}(\Omega)$}

In this subsection we state the final result for $\Re\sigma_{zz,s'}(\Omega)$. Essential steps are mentioned in the Appendix B. There we see that the same Eq.(\ref{sigma-xx-I}) and (\ref{sigma-xx-II}) 
hold for $\frac{\Re\sigma_{zz,s'}(\Omega)}{e^2/8\pi}$ but with $I_{s'}$ replaced by $L_{s'}$ with,

\bea
&& \hspace{-0.4cm} L^{1}_{s'}=\frac{k_{0}v_{z}}{2^{\frac{2}{J}}J v_{\perp}} \lp \frac{\Omega}{\Omega_{0}}\rp^{\frac{2-J}{J}} \biggl[ \frac{\sqrt{\pi}\Gamma(1+\frac{1}{J})}{2\Gamma(\frac{3}{2}+\frac{1}{J})}-\nonumber\\ 
&& \hspace{-0.4cm} \frac{1}{C'_{s'}}\lp\frac{2|\mu_{s'}|}{\Omega}-1\rp {_{2}F}_{1} \lb \frac{1}{2},-\frac{1}{J};\frac{3}{2};\frac{1}{C'^2_{s'}} \lp \frac{2|\mu_{s'}|}{\Omega}-1\rp^2 \rb \biggr],\nonumber\\
&& \hspace{-0.4cm} L^{2}_{s'}=\frac{k_{0}v_{z}}{2^{\frac{2}{J}}J v_{\perp}}\lp\frac{\Omega}{\Omega_{0}}\rp^{\frac{2-J}{J}} \frac{\sqrt{\pi} \Gamma(1+\frac{1}{J})}{\Gamma(\frac{3}{2}+\frac{1}{J})},\nonumber\\
&& \hspace{-0.4cm} L^{3}_{s'}= \frac{k_{0}v_{z}}{2^{\frac{2}{J}}J v_{\perp}C'_{s'}} \lp \frac{\Omega}{\Omega_{0}}\rp^{\frac{2-J}{J}}\times \nonumber \\
&& \hspace{-0.4cm} \biggl[\lp1+\frac{2|\mu_{s'}|}{\Omega}\rp {_{2}F}_{1} \lb \frac{1}{2},-\frac{1}{J};\frac{3}{2};\frac{1}{C'^2_{s'}}\lp 1+\frac{2|\mu_{s'}|}{\Omega}\rp^2 \rb + \nonumber\\
&& \hspace{-0.4cm} \lp 1-\frac{2|\mu_{s'}|}{\Omega}\rp {_{2}F}_{1} \lb \frac{1}{2},-\frac{1}{J};\frac{3}{2};\frac{1}{C'^2_{s'}}\lp 1-\frac{2|\mu_{s'}|}{\Omega}\rp^2 \rb \biggr],
\label{Ls}
\eea
where $\Omega_{0}=v_{\perp}k_{0}$ and ${_{2}F}_{1}$ is the Hypergeometric function defined as 
\be
{_{2}F}_{1}\lp a,b;c;z \rp = \sum^{\infty}_{n=0} \frac{(a)_{n}(b)_{n}}{(c)_{n}}\frac{z^{n}}{n!},
\ee
with $(q)_{n}$ the (rising) Pochhammer symbol, which is defined by
\bea
&& (q)_{n}=1 ~~~~~~~~~~~~~~~~~~~~~~~~~~~~~~~~~\text{for}~~~ n=0 \nonumber \\
&& ~~~~~~=q(q+1)...(q+n-1)~~~~~~~\text{for}~~~ n>0.
\eea
These formulas are central to the current article. In the next two sections we will use various special cases of these formulas to understand the implication of the inversion symmetry breaking, of tilting 
and of doping on the conductivity. For zero tilt we get,
\be
\hspace{-0.15cm} \frac{\Re\sigma_{zz,s'}(\Omega)}{e^2/8\pi}=L^2_{s'}=\frac{k_{0}v_{z}}{2^{\frac{2}{J}}J v_{\perp}} \lp\frac{\Omega}{\Omega_{0}}\rp^{\frac{2-J}{J}}\hspace{-0.1cm}\frac{\sqrt{\pi}\Gamma(1+\frac{1}{J})}{\Gamma(\frac{3}{2}+\frac{1}{J})}
\label{no-tilt}
\ee
for $\Omega>2\mu_{s'}$ and zero for $\Omega<2\mu_{s'}$, a known result \cite{Mele} in the limit of $Q_{0}=0$. Note that the displacement in momentum of the two Weyl nodes ($Q$) has dropped out of our 
clean limit calculations of the absorptive part of the conductivity. Of course, as is well known \cite{Burkov,Burkov2,Tiwari} it plays a crucial role in the DC Hall conductivity which is found to be 
proportional to $Q$ \cite{Burkov,Burkov2,Tiwari}. This parameter also enters prominently in other transport coefficients\cite{Ferreiros,Saha}.

\section{Conductivities with inversion symmetry}
\label{IV}

\begin{figure}[H]
\centering
\includegraphics[width=2.5in,height=3.0in, angle=270]{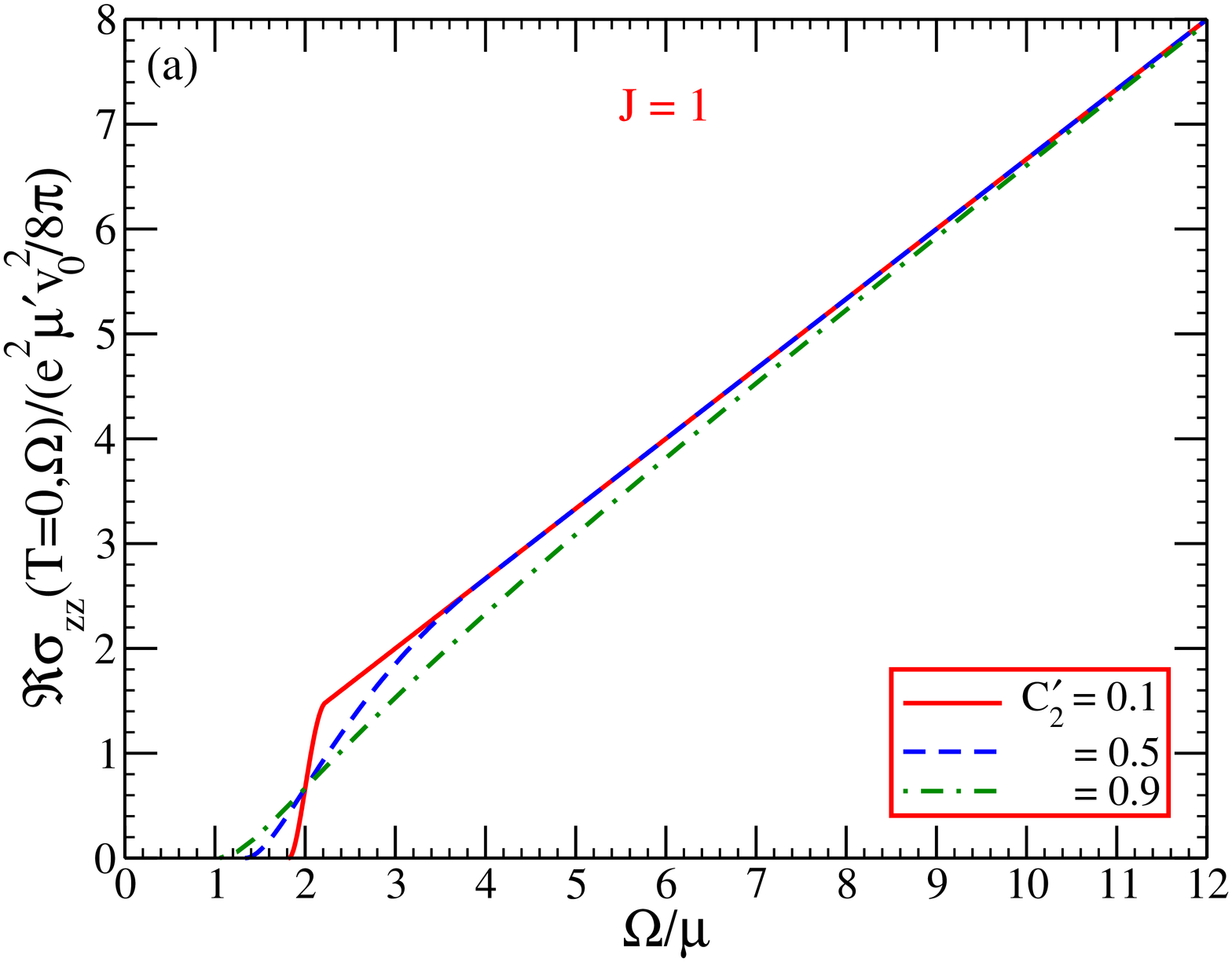}
\includegraphics[width=2.5in,height=3.0in, angle=270]{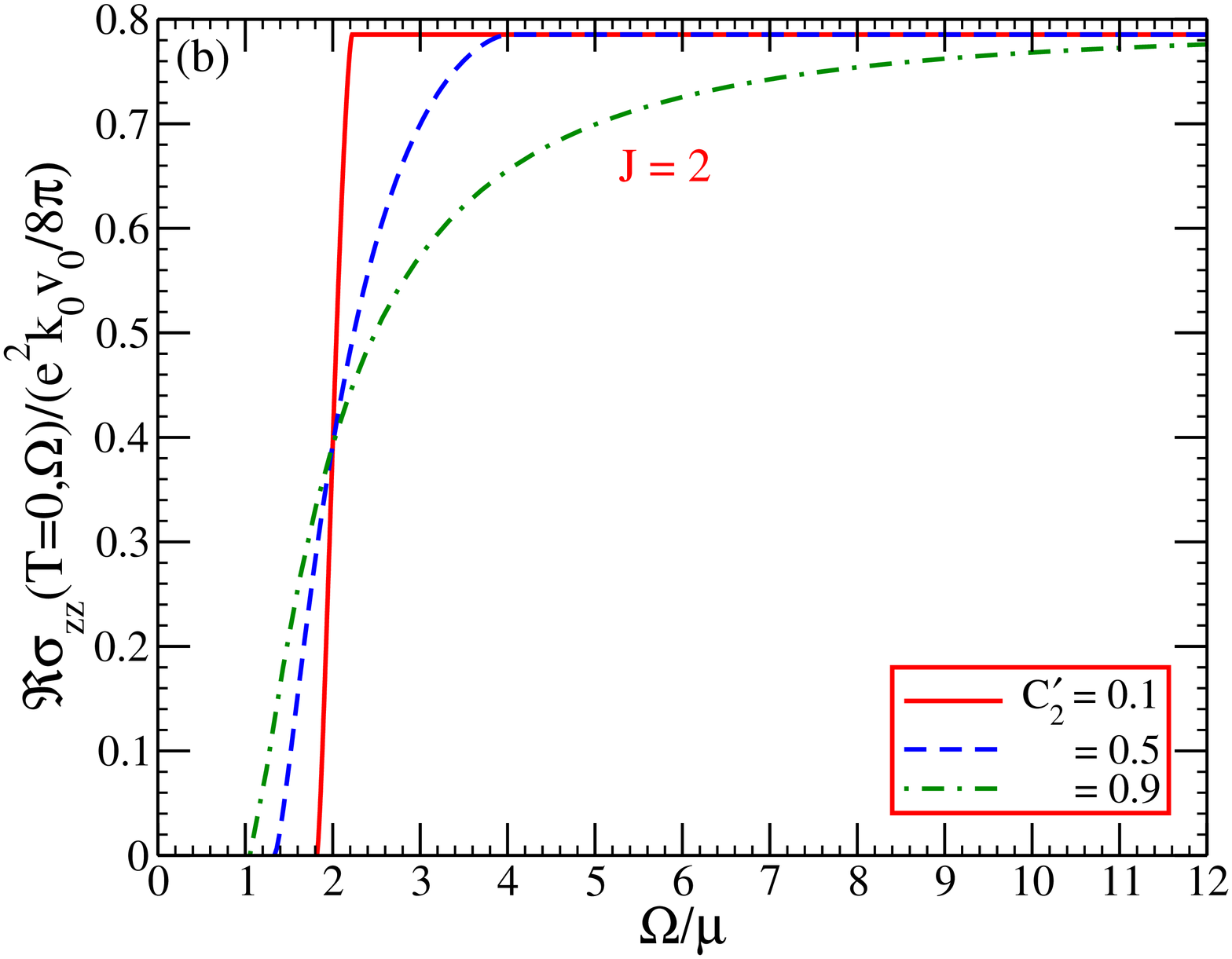}
\caption{(Color online) The real part of the dynamic optical conductivity in $z$-direction $\Re{\sigma_{zz}}(T=0,\Omega)$ at temperature $T=0$ in units of $e^2\mu'v^2_{0}/8\pi$(for $J=1$) and 
$e^2k_{0}v_{0}/8\pi$(for $J=2$) as a function of photon energy $\Omega$ normalized to the chemical potential $\mu$. Top frame (a) corresponds to $J=1$ while the bottom frame (b) is for $J=2$. They 
compare results for different values of tilt namely $C'_{2}=0.1$ (solid red), $C'_{2}=0.5$ (dash blue) and $C'_{2}=0.9$ (dash-dot green) all type I. No inversion symmetry breaking is included ($Q_{0}=0$). 
Winding number has a large effect on this quantity including the result of tilting. }
\label{Fig2}
\end{figure}

\begin{figure}[H]
\centering
\includegraphics[width=2.5in,height=3.0in, angle=270]{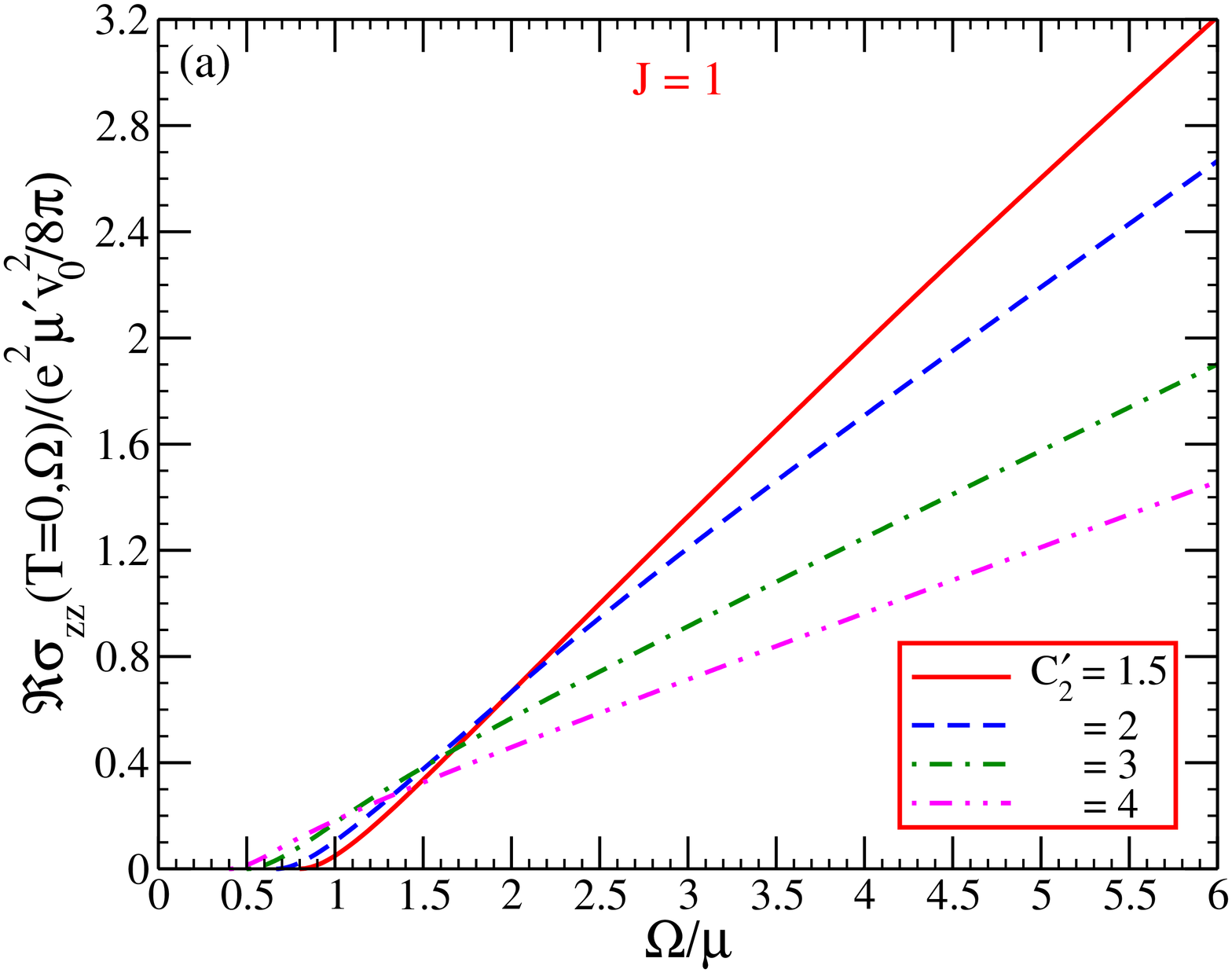}
\includegraphics[width=2.5in,height=3.0in, angle=270]{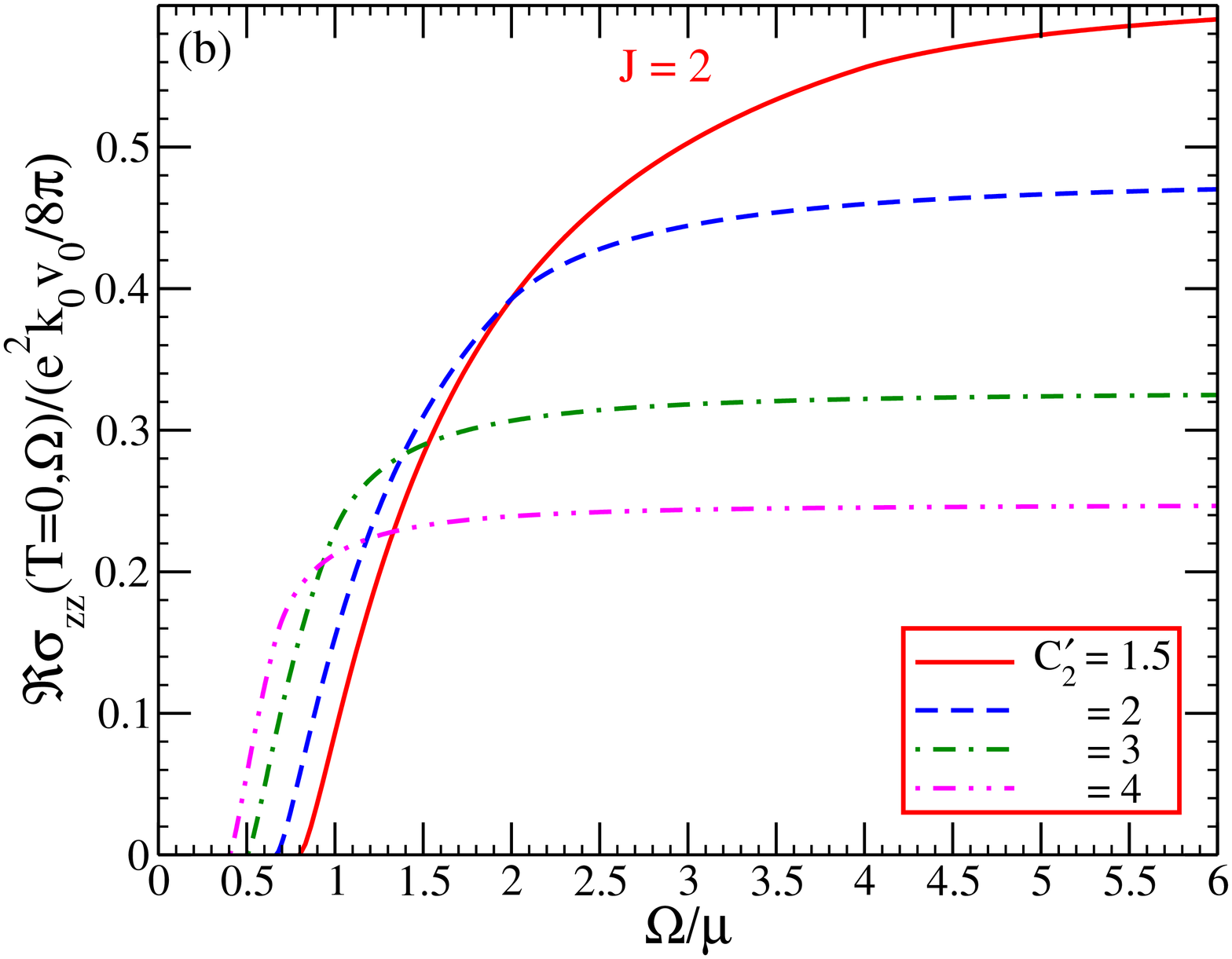}
\caption{(Color online) The real part of the dynamic optical conductivity in $z$-direction $\Re{\sigma_{zz}}(T=0,\Omega)$ at temperature $T=0$ in units of $e^2\mu'v^2_{0}/8\pi$ (for $J=1$) and 
$e^2k_{0}v_{0}/8\pi$ (for $J=2$) as a function of photon energy $\Omega$ normalized to the chemical potential $\mu$. Top frame (a) corresponds to $J=1$ while the bottom frame (b) is for $J=2$. They 
compare results for different values of tilt namely $C'_{2}=1.5$ (solid red), $C'_{2}=2$ (dash blue), $C'_{2}=3$ (dash-dot green) and $C'_{2}=4$ (dash double dot magenta) all type II. No inversion 
symmetry breaking is included ($Q_{0}=0$). Winding number has a large effect on this quantity including the result of tilting.} 
\label{Fig3}
\end{figure}

We begin this section with the case when $Q_{0}=0$ (centrosymmetric). We need only consider $\Re\sigma_{zz,s'}(\Omega)$ for $J=1$ as $\Re\sigma_{xx,s'}(\Omega)$ has been discussed before\cite{Mukherjee1}.
For $Q_{0}=0$ the $L$-functions defined in Eq.(\ref{Ls}) becomes independent of $s'$ as $\mu_{s'}$ is now replaced by $\mu$ and $C'_{s'}$ is replaced by $C'_{2}$ because we assume that the tilt has the 
same magnitude on both nodes. Tilt inversion symmetry guarantees that the conductivity only depends on its absolute value. The total contribution to the conductivity is therefore twice the amount in 
Eq.(\ref{Ls}). This also modifies the $\Omega$ dependent prefactor in all the $L$-functions to $\frac{2k_{0}v_{z}}{2^{\frac{2}{J}}J v_{\perp}}\lp \frac{\Omega}{\Omega_{0}}\rp^{\frac{2-J}{J}}$ which can 
be equivalently written as $\mu'v^2_{0}\tOmega$ ($J=1$) and $k_{0}v_{0}$ ($J=2$). This also change $\tOmega^{s'}_{L}$ to $\tOmega_{L}=\frac{2}{1+C'_{2}}$ and $\tOmega^{s'}_{U}$ to 
$\tOmega_{U}=\frac{2}{|1-C'_{2}|}$. With all this in mind we plot  $\Re\sigma_{zz,s'}(\Omega)$ (in appropriate unit) for type I in Fig.[\ref{Fig2}] and type II in Fig.[\ref{Fig3}] for both 
$J=1~(\text{top frames}),2~(\text{bottom frames})$.
In Fig.[\ref{Fig2}] we consider three representative values of tilt, $C'_{2}=0.1$ (solid red), $C'_{2}=0.5$ (dash blue) and $C'_{2}=0.9$ (dash-dot green). The solid red curve includes only a small value 
of tilt and is very close to the no tilt result. In this case $\Re\sigma_{zz,s'}(\Omega)$ is zero till $\Omega=2\mu$ where for $J=1$ it discontinuously jumps up to merge with the well known linear in 
photon energy ($\Omega$) law associated with the interband background absorption of a Dirac cone \cite{Nicol, Tabert, Carbotte}. The optical spectral weight in the interband background below $2\mu$ is 
transferred, due to Pauli blocking, to the Drude interband contribution. In the clean limit used here, this is a delta function at $\Omega=0$ not shown in our plot. While the sharp jump at $\Omega=2\mu$ 
of the no tilt case remains clearly seen in the red curve of Fig.[\ref{Fig2}a] this is not so for the other two curves dash blue ($C'_{2}=0.5$) and dash-dot green ($C'_{2}=0.9$). These are qualitatively 
better described as a roughly quasilinear rise out of zero at $\Omega=2\mu/(1+C'_{2})$ and merging with the zero tilt case at $\Omega=2\mu/(1-C'_{2})$. In the units used, the slope of the interband background
in the untilted case is $2/3$, and the conductivity at $\Omega=2\mu$ is equal to $4/3$ but for the tilted case it is instead half this value namely $2/3$. The quasilinear law in the photon energy interval 
$2\mu/(1+C'_{2})$ to $2\mu/(1-C'_{2})$ passes exactly through the half way mark of the sharp absorption edge of the untilted case. This follows for type I from Eq.(\ref{Ls}) which gives,
\be
\hspace{-0.15cm} \frac{\Re\sigma_{zz,s'}(2\mu)}{e^2/8\pi}=L^1_{s'}(2\mu)=\frac{k_{0}v_{z}}{2^{\frac{2}{J}}J v_{\perp}} \lp\frac{2\mu}{\Omega_{0}}\rp^{\frac{2-J}{J}}\hspace{-0.1cm}\frac{\sqrt{\pi}\Gamma(1+\frac{1}{J})}{2\Gamma(\frac{3}{2}+\frac{1}{J})}.
\label{no-tilt1}
\ee
This is exactly half the value of the no-tilt conductivity at the absorption edge given by Eq.(\ref{no-tilt}) evaluated at $\Omega=2\mu$. It is completely independent of the tilt so that all the curves 
in Fig.[\ref{Fig2}a] pass through this same point $\frac{\Re\sigma_{zz}(2\mu)}{e^2/8\pi}=\frac{2}{3}v^2_{0}\mu'$. Similar results hold in the case of winding number $J=2$ (multi Weyl). The interband 
optical background is no longer linear in photon energy for no tilt. Instead it is constant \cite{Mele} as in graphene\cite{Gusynin}
\be
\frac{\Re\sigma_{zz}(\Omega)}{e^2/8\pi}=\frac{\pi}{4} k_{0}v_{0}
\ee
as we see in Fig.[\ref{Fig2}b] where all curves pass through $\pi/8$ in the units used for the conductivity. As compared with the top frame the sharp jump at $\Omega=2\mu$ of the no tilt case 
remains more discernible with increasing tilt.

In Fig.[\ref{Fig3}] we present a series of results for type II Weyl ($C'_{2}>1$). A first observation is that as long as $C'_{2}<2$ Eq.(\ref{no-tilt1}) still gives the value of the conductivity at 
$\Omega=2\mu$. However for $C'_{2}>2$ 
\bea
&& \frac{\Re\sigma_{zz,s'}(2\mu)}{e^2/8\pi}=L^3_{s'}(2\mu)\nonumber \\
&& =\frac{2k_{0}v_{z}}{2^{\frac{2}{J}}J v_{\perp}C'_{2}} \lp\frac{2\mu}{\Omega_{0}}\rp^{\frac{2-J}{J}}{_{2}F}_{1} \lb \frac{1}{2},-\frac{1}{J};\frac{3}{2};\lp\frac{2}{C'_{2}}\rp^2\rb,
\label{no-tilt2}
\eea
with
\bea
&& {_{2}F}_{1} \lb \frac{1}{2},-\frac{1}{J};\frac{3}{2};x^2 \rb = 1-\frac{x^2}{3}, ~~~\text{for}~~~J=1\nonumber\\
&& =\frac{\sqrt{1-x^2}}{2}+\frac{\arcsin x}{2x}, ~~~\text{for}~~~J=2,
\label{HGF}
\eea
\noindent
which works out to 
\bea
&& \hspace{-0.2cm}\frac{\Re\sigma_{zz,s'}(2\mu)}{e^2/8\pi}=v^2_{0}\mu' \frac{1}{C'_{2}} \lb 1-\frac{1}{3} \lp\frac{2}{C'_{2}}\rp^2 \rb(\text{for}~J=1)\\
&& \hspace{-0.2cm}=\frac{k_{0}v_{0}}{4} \lb \frac{1}{C'_{2}}\sqrt{1-\lp\frac{2}{C'_{2}}\rp^2}+\frac{1}{2}\arcsin \lp\frac{2}{C'_{2}}\rp\rb(\text{for}~J=2)\nonumber,
\label{tiltedII}
\eea
both now depend on the tilt as is verified in the numerical work of Fig.[\ref{Fig3}]. We present results for $C'_{2}=1.5$ (solid red), $C'_{2}=2$ (dash blue), $C'_{2}=3$ (dash-dot green) and $C'_{2}=4$ 
(dash double dot magenta). As $C'_{2}$ is increased the curves start at smaller values of photon energy, eventually crossing each other and at the larger values of $\Omega$ shown, the slope of the 
quasilinear behavior in the top frame ($J=1$) decreases while in the lower frame ($J=2$) they flatten out at ever smaller values. For $\Omega>\frac{2\mu}{C'_{2}-1}$ and $C'_{2}$ going to infinity 
Eq.(\ref{Ls}) gives,
\bea
&& \frac{\Re\sigma_{zz,s'}(\Omega)}{e^2/8\pi}=L^3_{s'}(\Omega)\nonumber \\
&& \cong \frac{k_{0}v_{z}}{2^{\frac{2}{J}-1}J v_{\perp}C'_{2}} \lp\frac{\Omega}{\Omega_{0}}\rp^{\frac{2-J}{J}}{_{2}F}_{1} \lb \frac{1}{2},-\frac{1}{J};\frac{3}{2};\frac{1}{C'^{2}_{2}}\rb
\eea
which is to leading order
\bea
&& \hspace{-0.6cm}\cong  \frac{k_{0}v_{z}}{2^{\frac{2}{J}-1}J v_{\perp}C'_{2}} \lp\frac{\Omega}{\Omega_{0}}\rp^{\frac{2-J}{J}} ~~(\text{for} J=1)\nonumber\\
&& \hspace{-0.6cm}\cong  \frac{k_{0}v_{z}}{2^{\frac{2}{J}-1}J v_{\perp}C'_{2}} \lp\frac{\Omega}{\Omega_{0}}\rp^{\frac{2-J}{J}}~~(\text{for} J=2).
\eea
There is a $1/C'_{2}$ decay factor as the tilt gets very large. More specifically
\be
\frac{\Re\sigma_{zz,s'}(\Omega)}{e^2/8\pi}=\frac{v^{2}_{0}\Omega}{2v_{z}C'_{2}}~~(\text{for} J=1)
\label{J1}
\ee
and
\be
\frac{\Re\sigma_{zz,s'}(\Omega)}{e^2/8\pi}=\frac{k_{0}v_{0}}{2C'_{2}}~~(\text{for} J=2).
\label{J2}
\ee
Consequently the slope of the linear in $\Omega$ law of Eq.(\ref{J1}) decays as the inverse of the tilt parameter for $J=1$ and the magnitude of the constant value in Eq.(\ref{J2}) also decays as one over 
the tilt. It is interesting to compare these relations with the case of $\Re\sigma_{xx,s'}(\Omega)$ given in Eq.(\ref{sigma-xx-II})
\be
\frac{\Re\sigma_{xx,s'}(\Omega)}{\mu'e^2/8\pi}=I^{3}_{s'}\cong \frac{J\Omega}{4C'_{2}\mu}, 
\ee
in the limit of interest here. This differs from Eq.(\ref{J1}) by a factor of $2v^{2}_{0}$ for the case $J=1$.

\section{AC conductivity when $Q_{0}\ne0$}
\label{V}

In this section we describe the effect of the finite $Q_{0}$ on both the AC conductivities $\Re{\sigma_{xx}}(T=0,\Omega)$ and $\Re{\sigma_{zz}}(T=0,\Omega)$. Here we make the same assumption as before 
that both the nodes have the same absolute value of tilt. The tilt direction makes no difference. 

I. For $0<C'_{2}<\tQ_{0}$ 
\bea
&& \frac{\Re\sigma_{xx}(\Omega)}{\mu' e^2/8\pi}=0,~~\text{where}~~~~ \tOmega<\tOmega^{-}_{L}\nonumber\\
&&\hspace{1.5cm}=I^{1}_{-},~~\text{where}~~~~ \tOmega^{-}_{U}>\tOmega>\tOmega^{-}_{L}\nonumber\\
&&\hspace{1.5cm}=I^{2}_{-},~~\text{where}~~~~ \tOmega^{+}_{L}>\tOmega>\tOmega^{-}_{U}\nonumber\\
&&\hspace{1.5cm}=(I^{1}_{+}+I^{2}_{-}),~~\text{where}~~~~ \tOmega^{+}_{U}>\tOmega>\tOmega^{+}_{L}\nonumber\\
&&\hspace{1.5cm}=(I^{2}_{+}+I^{2}_{-}), ~~\text{where}~~~~ \tOmega>\tOmega^{+}_{U}
\label{sigma-xx-I1}
\eea

II. For $\tQ_{0}<C'_{2}<1$ 
\bea
&& \frac{\Re\sigma_{xx}(\Omega)}{\mu' e^2/8\pi}=0,~~\text{where}~~~~ \tOmega<\tOmega^{-}_{L}\nonumber\\
&&\hspace{1.5cm}=I^{1}_{-},~~\text{where}~~~~ \tOmega^{+}_{L}>\tOmega>\tOmega^{-}_{L}\nonumber\\
&&\hspace{1.5cm}=(I^{1}_{+}+I^{1}_{-}),~~\text{where}~~~~ \tOmega^{-}_{U}>\tOmega>\tOmega^{+}_{L}\nonumber\\
&&\hspace{1.5cm}=(I^{1}_{+}+I^{2}_{-}),~~\text{where}~~~~ \tOmega^{+}_{U}>\tOmega>\tOmega^{-}_{U}\nonumber\\
&&\hspace{1.5cm}=(I^{2}_{+}+I^{2}_{-}),~~\text{where}~~~~ \tOmega>\tOmega^{+}_{U}
\label{sigma-xx-I2}
\eea

III. For $C'_{2}>1$ but $C'_{2}\tQ_{0}<1$
\bea
&& \frac{\Re\sigma_{xx}(\Omega)}{\mu' e^2/8\pi}=0,~~\text{where}~~~~ \tOmega<\tOmega^{-}_{L}\nonumber\\
&&\hspace{1.5cm}=I^{1}_{-},~~\text{where}~~~~ \tOmega^{+}_{L}>\tOmega>\tOmega^{-}_{L}\nonumber\\
&&\hspace{1.5cm}=(I^{1}_{+}+I^{1}_{-}),~~\text{where}~~~~ \tOmega^{-}_{U}>\tOmega>\tOmega^{+}_{L}\nonumber\\
&&\hspace{1.5cm}=(I^{1}_{+}+I^{3}_{-}),~~\text{where}~~ \tOmega^{+}_{U}>\tOmega>\tOmega^{-}_{U}\nonumber\\
&&\hspace{1.5cm}=(I^{3}_{+}+I^{3}_{-}),~~\text{where}~~~~ \tOmega>\tOmega^{+}_{U}
\label{sigma-xx-II1}
\eea

IV. For $C'_{2}>1$ and $C'_{2}\tQ_{0}>1$

\bea
&& \frac{\Re\sigma_{xx}(\Omega)}{\mu' e^2/8\pi}=0,~~\text{where}~~~~ \tOmega<\tOmega^{-}_{L}\nonumber\\
&&\hspace{1.5cm}=I^{1}_{-},~~\text{where}~~~~ \tOmega^{-}_{U}>\tOmega>\tOmega^{-}_{L}\nonumber\\
&&\hspace{1.5cm}=I^{3}_{-},~~\text{where}~~~~ \tOmega^{+}_{L}>\tOmega>\tOmega^{-}_{U}\nonumber\\
&&\hspace{1.5cm}=(I^{1}_{+}+I^{3}_{-}),~~\text{where}~~ \tOmega^{+}_{U}>\tOmega>\tOmega^{+}_{L}\nonumber\\
&&\hspace{1.5cm}=(I^{3}_{+}+I^{3}_{-}),~~\text{where}~~~~ \tOmega>\tOmega^{+}_{U}.
\label{sigma-xx-II2}
\eea
For $\Re{\sigma_{zz}}(T=0,\Omega)$ we only need to replace $I_{s'}$ in Eq.(\ref{sigma-xx-I1}) to (\ref{sigma-xx-II2}) by the $L_{s'}$ of Eq.(\ref{Ls}) and drop the $\mu'$ on the left hand side of these 
equations.

\begin{figure}[H]
\centering
\includegraphics[width=2.5in,height=3.0in, angle=270]{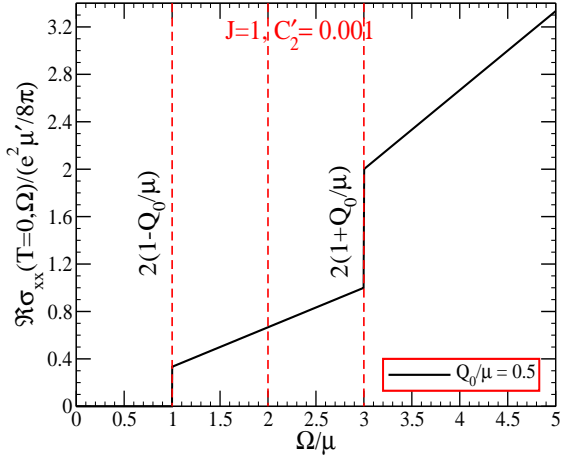}
\caption{(Color online)The real part of the dynamic optical conductivity perpendicular to the $z$-axis $\Re{\sigma_{xx}}(T=0,\Omega)$ at zero temperature in units of $e^2\mu'/8\pi$ as a function of photon 
energy $\Omega$ normalized to the chemical potential $\mu$. This plot includes broken inversion symmetry with $Q_{0}/\mu=0.5$. The winding number is $J=1$ and a small tilt $C'_{2}=0.001$ is included but
has negligible effect on the plot.} 
\label{Fig4}
\end{figure}
\begin{figure}[H]
\centering
\includegraphics[width=2.5in,height=3.0in, angle=270]{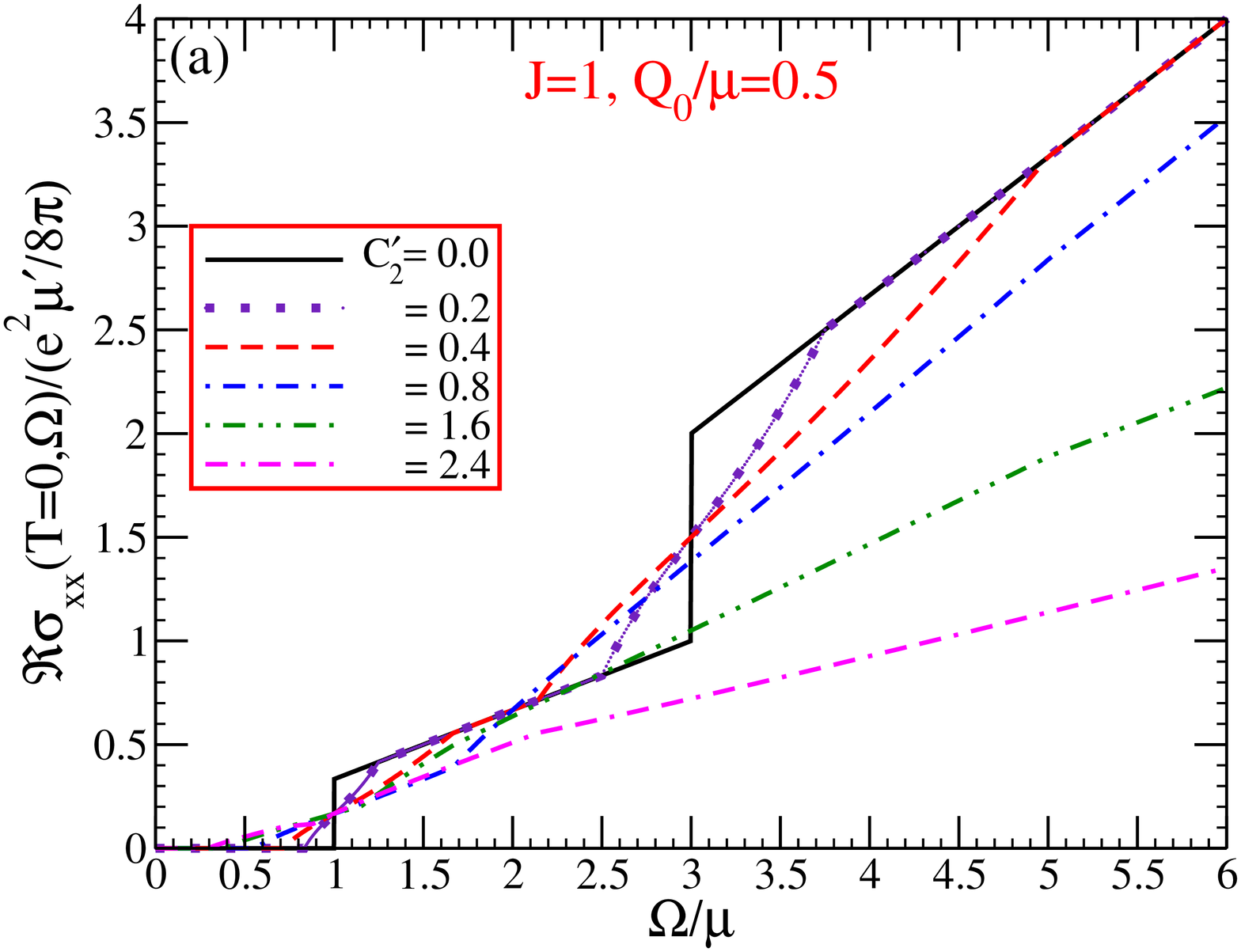}
\includegraphics[width=2.5in,height=3.0in, angle=270]{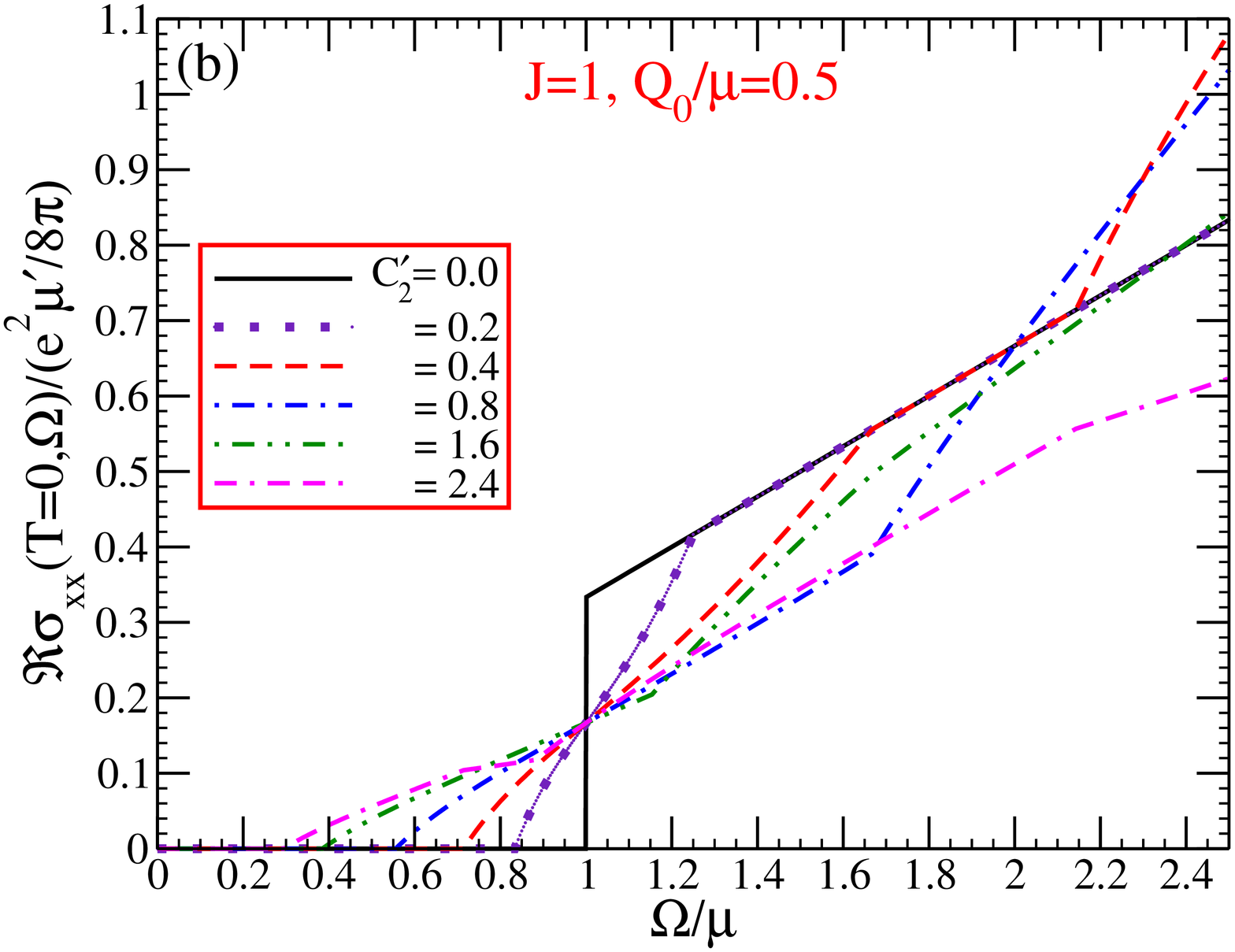}
\caption{(Color online) The real part of the dynamic optical conductivity perpendicular to the $z$-axis $\Re{\sigma_{xx}}(T=0,\Omega)$ at zero temperature in units of $e^2\mu'/8\pi$ as a function of photon 
energy $\Omega$ normalized to the chemical potential $\mu$. The top frame (a) shows results up to $\Omega/\mu=6$ while the bottom frame (b) is an expanded version of the low energy region to $\Omega/\mu=2.5$
only. In both frames the winding number is $J=1$ and the inversion symmetry breaking parameter has been set to $Q_{0}/\mu=0.5$. Six values of tilt are shown (solid black) $C'_{2}=0$, (dot indigo) $C'_{2}=0.2$, 
(dash red) $C'_{2}=0.4$, and (dash-dot blue) $C'_{2}=0.8$ (type I), (dash double dot green) $C'_{2}=1.6$, (dot double dash magenta) $C'_{2}=2.4$. } 
\label{Fig5}
\end{figure}
\begin{figure}[H]
\centering
\includegraphics[width=2.5in,height=3.0in, angle=270]{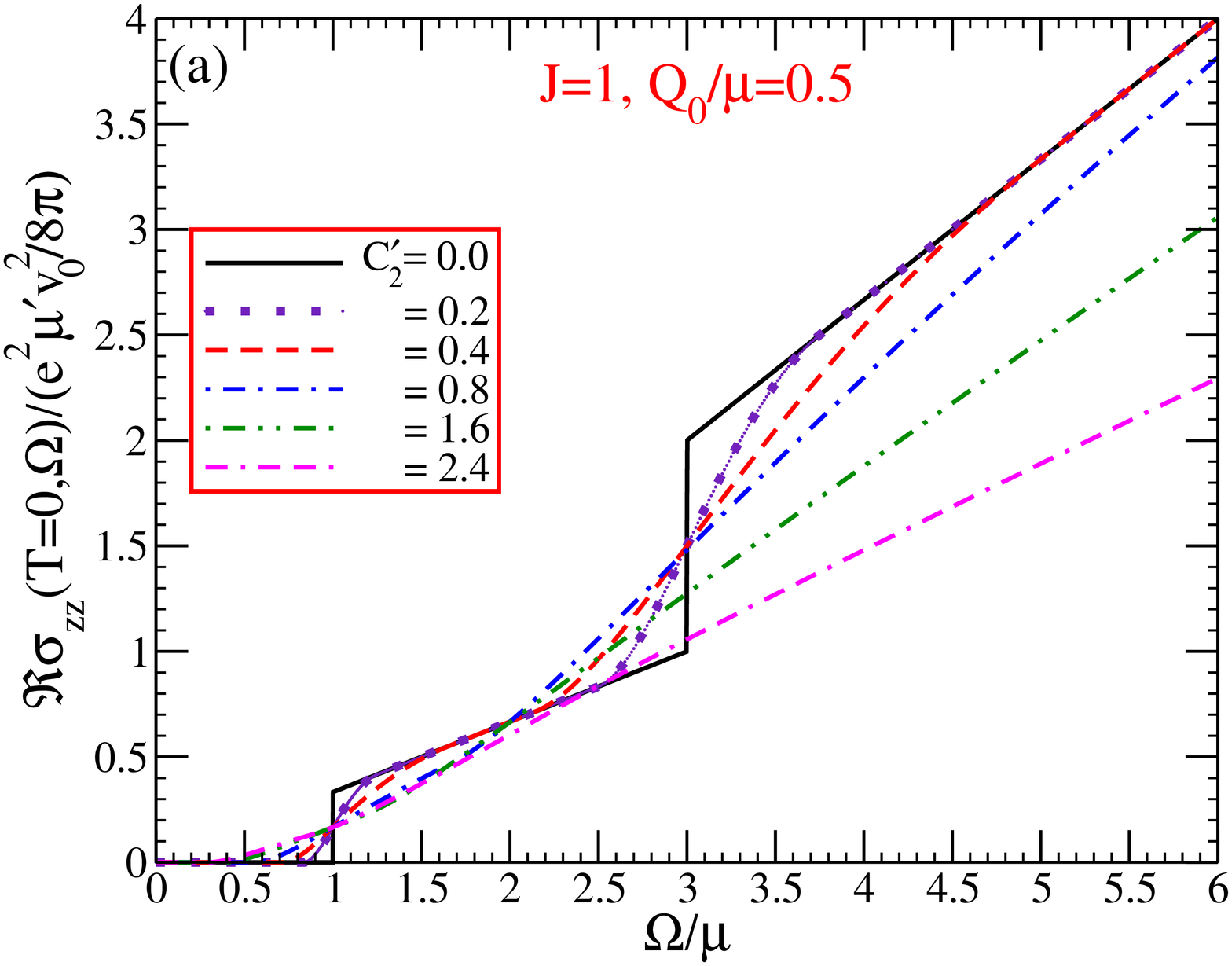}
\includegraphics[width=2.5in,height=3.0in, angle=270]{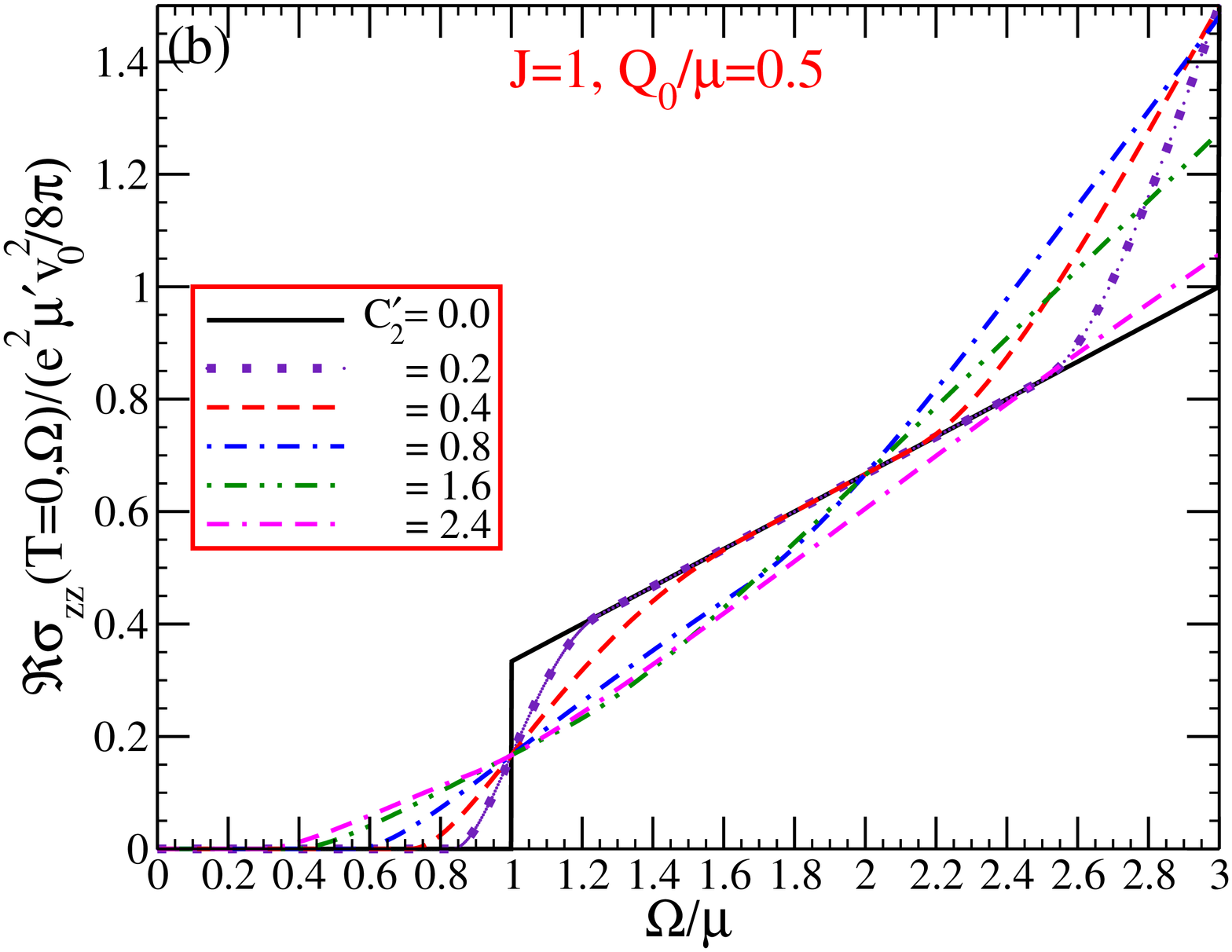}
\caption{(Color online) The real part of the dynamic optical conductivity along the $z$-axis $\Re{\sigma_{zz}}(T=0,\Omega)$ at zero temperature in units of $e^2\mu'v^2_{0}/8\pi$ as a function of photon 
energy $\Omega$ normalized to the chemical potential $\mu$. This plot is to be compared with Fig.[\ref{Fig5}]. The top frame (a) shows results up to $\Omega/\mu=6$ while the bottom frame (b) is an expanded 
version of the low energy region to $\Omega/\mu=3$ only. In both frames the winding number is $J=1$ and the inversion symmetry breaking parameter has been set to $Q_{0}/\mu=0.5$. Six values of tilt are 
shown (solid black) $C'_{2}=0$, (dot indigo) $C'_{2}=0.2$, (dash red) $C'_{2}=0.4$, and (dash-dot blue) $C'_{2}=0.8$ (type I), (dash double dot green) $C'_{2}=1.6$, (dot double dash magenta) $C'_{2}=2.4$.} 
\label{Fig6}
\end{figure}

In Fig.[\ref{Fig4}] we plot $\Re{\sigma_{xx}}(T=0,\Omega)$ in units of $e^2\mu'/8\pi$ as a function of photon energy $\Omega$ normalized to the chemical potential $\mu$ for $Q_{0}=0.5$ and $C'_{2}$ close 
to zero. The plot follows directly from Eq.(\ref{sigma-xx-I1}) and reproduces a similar plot in Ref.[\onlinecite{Nicol}]. The contribution of the negative chirality node has the effective chemical potential 
$\mu(1-Q_{0}/\mu)=\mu/2$ jumps from zero to a value of $1/3$ at $\Omega=\mu$ in the units of Fig.[\ref{Fig4}]. For the positive chirality 
node the effective chemical potential is instead $3\mu$ and the second jump in the conductivity has magnitude one. The slope of the first straight line between $\Omega/\mu$ equal to 1 and 3 is $1/3$ and 
involves only the negative chirality node while in the second interval above $\Omega/\mu=3$ it is twice this value and involves both nodes. The double jump behavior seen in this figure is the hallmarks 
of broken inversion symmetry i.e. a finite value of $Q_{0}$. When a finite tilt is included in the calculation it will modify the contribution from each node as shown in Fig.[\ref{Fig5}] and Fig.[\ref{Fig6}] 
for $\Re\sigma_{xx}(\Omega)$ and $\Re\sigma_{zz}(\Omega)$ respectively. Both are for $J=1$ and $Q_{0}/\mu=0.5$. The solid black curve in Fig.[\ref{Fig5}] is the same as shown in Fig.[\ref{Fig4}] with a 
first step at $\Omega/\mu=1$ and a second at $\Omega/\mu=3$. As $C'_{2}$ is increased dot indigo $C'_{2}=0.2$, dash red $C'_{2}=0.4$, dash-dot blue $C'_{2}=0.8$, dash double dot green $C'_{2}=1.6$ and 
dot double dash magenta $C'_{2}=2.4$ the steps progressively lose their integrity as the optical spectral weight is transferred from the region above twice the effective chemical potential of a given 
node to the region below this photon energy. The amount of transfer and the interval over which this transfer occurs increase as $C'_{2}$ is increased out of zero. As is seen in the lower frame of 
Fig.[\ref{Fig5}] where only the region $\Omega/\mu$ below 2.5 is shown on an expanded scale for the $C'_{2}=0.2$ (dot indigo) and $C'_{2}=0.4$ (dash red) the redistribution of optical spectral weight due 
to the tilt for the negative chirality node ends at $\Omega/\mu=1.25$ and 1.66 respectively, both are below the minimum photon energy for which the positive chirality node contributes which are 2.5 and 
2.14 respectively. In both instances we see a linear region with slope exactly equal to its no tilt value representing the photon energy interval over which only the negative chirality node contributes 
even though the tilt has spread out the optical spectral weight coming from the positive chirality node. In this region the contribution from the negative chirality node has also recovered its zero tilt 
slope. In all other curves this no longer happens and no perfectly linear region remains. 
\noindent
Note also that below $\Omega/\mu=1$ the curves for $C'_{2}=0.2$ and 0.4 show concave downward behavior which changes to concave upward as we go through $\Omega/\mu=1$. These effects are small and when 
ignored the underlying behavior is quasilinear.

The results for $\Re{\sigma_{zz}}(T=0,\Omega)$ in Fig.[\ref{Fig6}] has much the same qualitative behavior as for those in Fig.[\ref{Fig5}] but there are significant quantitative changes. One difference
is that the curves for $C'_{2}=0.2$ and 0.4 now show concave upward behavior below $\Omega/\mu=1$ which shift to concave downward above this energy. The high energy behavior is also quantitatively different.
When there is no tilt $\Re{\sigma_{zz}}(T=0,\Omega)$ is the same as $\Re{\sigma_{xx}}(T=0,\Omega)$ except for an extra factor of $v^2_{0}$ and this factor is one in the isotropic Weyl case ($J=1$). For 
type I, $\Re{\sigma_{xx}}(T=0,\Omega)$ reduces to $\frac{e^2}{24\pi}\frac{\Omega}{v_{z}}$ consistent with our previous results \cite{Nicol} once an $\hbar^2$ is restored. For type II ($C'_{2}>1$) we get 
for large values of $\Omega$,
\be
\Re{\sigma_{xx}}(T=0,\Omega)=\frac{e^2}{8\pi}\frac{\Omega}{v_{z}}\frac{1}{4C'_{2}}\lb 1+\frac{1}{3C'^{2}_{2}}\rb ~~~\text{and}
\ee
\be
\hspace{-1.4cm}\Re{\sigma_{zz}}(T=0,\Omega)=\frac{e^2}{8\pi}v^2_{0}\frac{\Omega}{v_{z}}\frac{1}{2C'_{2}}\lb 1-\frac{1}{3C'^{2}_{2}}\rb.
\ee
\begin{figure}[H]
\centering
\includegraphics[width=2.5in,height=3.0in, angle=270]{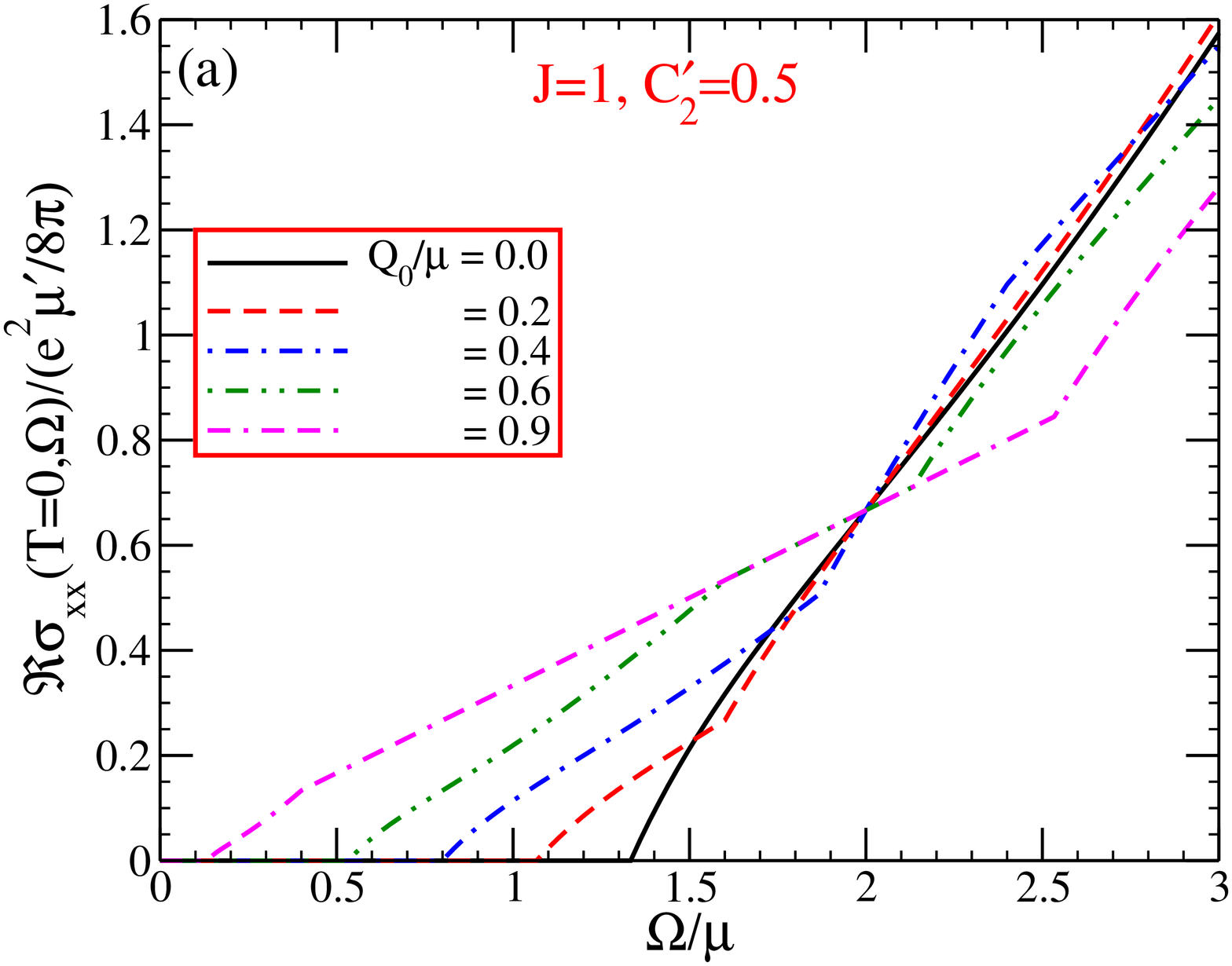}
\includegraphics[width=2.5in,height=3.0in, angle=270]{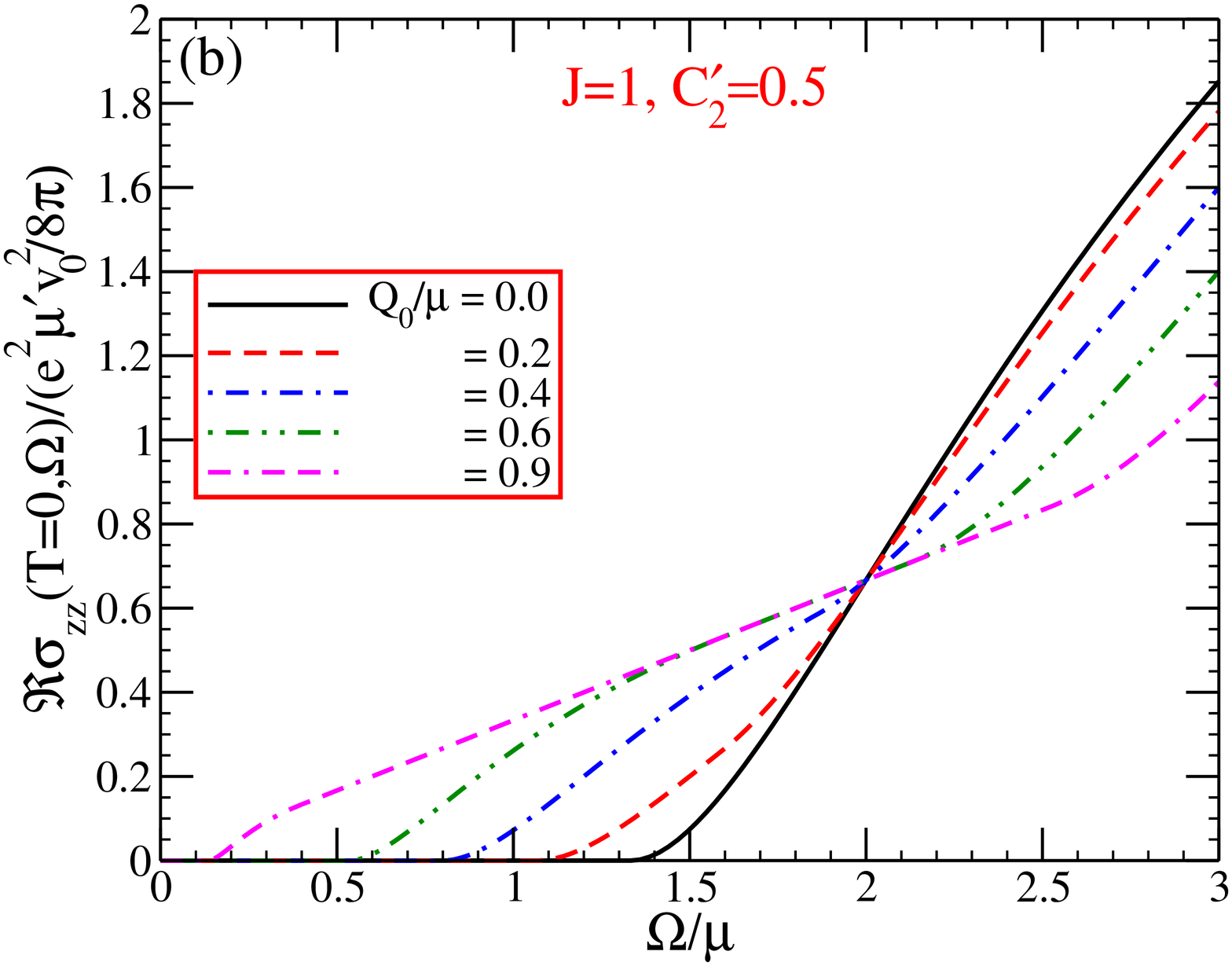}
\caption{(Color online)The real part of the dynamic optical conductivity perpendicular to the $z$-axis (frame (a)) $\Re{\sigma_{xx}}(T=0,\Omega)$ at zero temperature in units of $e^2\mu'/8\pi$ as a 
function of photon energy (normalized to the chemical potential $\mu$) up to $\Omega/\mu=3$. The tilt is $C'_{2}=0.5$ (type I). Five values are chosen for the inversion symmetry breaking parameter 
$Q_{0}/\mu$ namely $Q_{0}/\mu=0$ (solid black), $Q_{0}/\mu=0.2$ (dash red), $Q_{0}/\mu=0.4$ (dash-dot blue), $Q_{0}/\mu=0.6$ (dash double dot green) and $Q_{0}/\mu=0.9$ (dot double dash magenta). Frame 
(b) is the real part of the dynamical optical conductivity along the $z$-axis $\Re{\sigma_{zz}}(T=0,\Omega)$ at zero temperature in units of $e^2\mu'v^2_{0}/8\pi$ as a function of photon energy 
(normalized to the chemical potential $\mu$) up to $\Omega/\mu=3$.} 
\label{Fig7}
\end{figure}
\noindent
which shows differences between $zz$ and $xx$ conductivities beyond a factor of $v^2_{0}$.

\begin{figure}[H]
\centering
\includegraphics[width=2.5in,height=3.0in, angle=270]{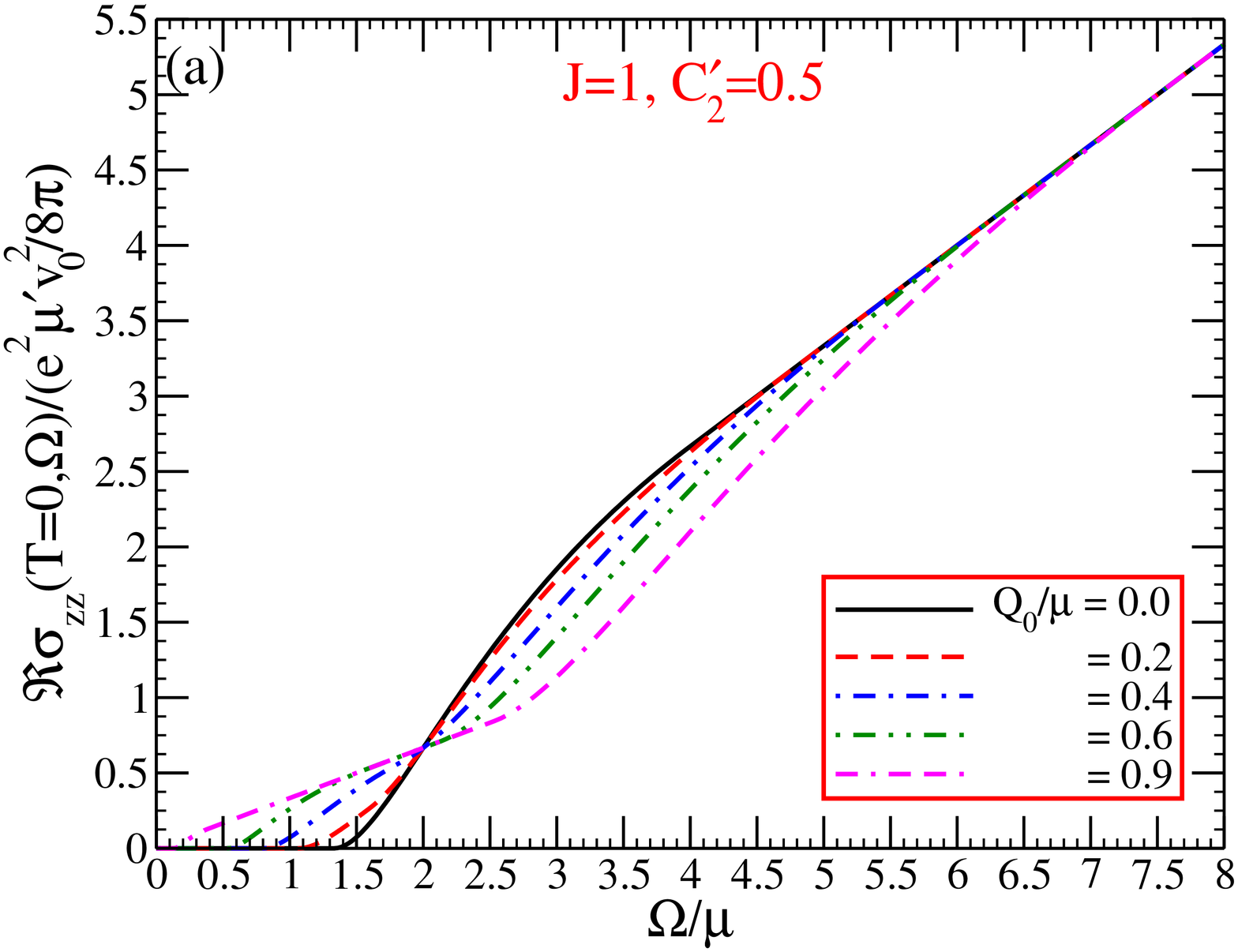}
\includegraphics[width=2.5in,height=3.0in, angle=270]{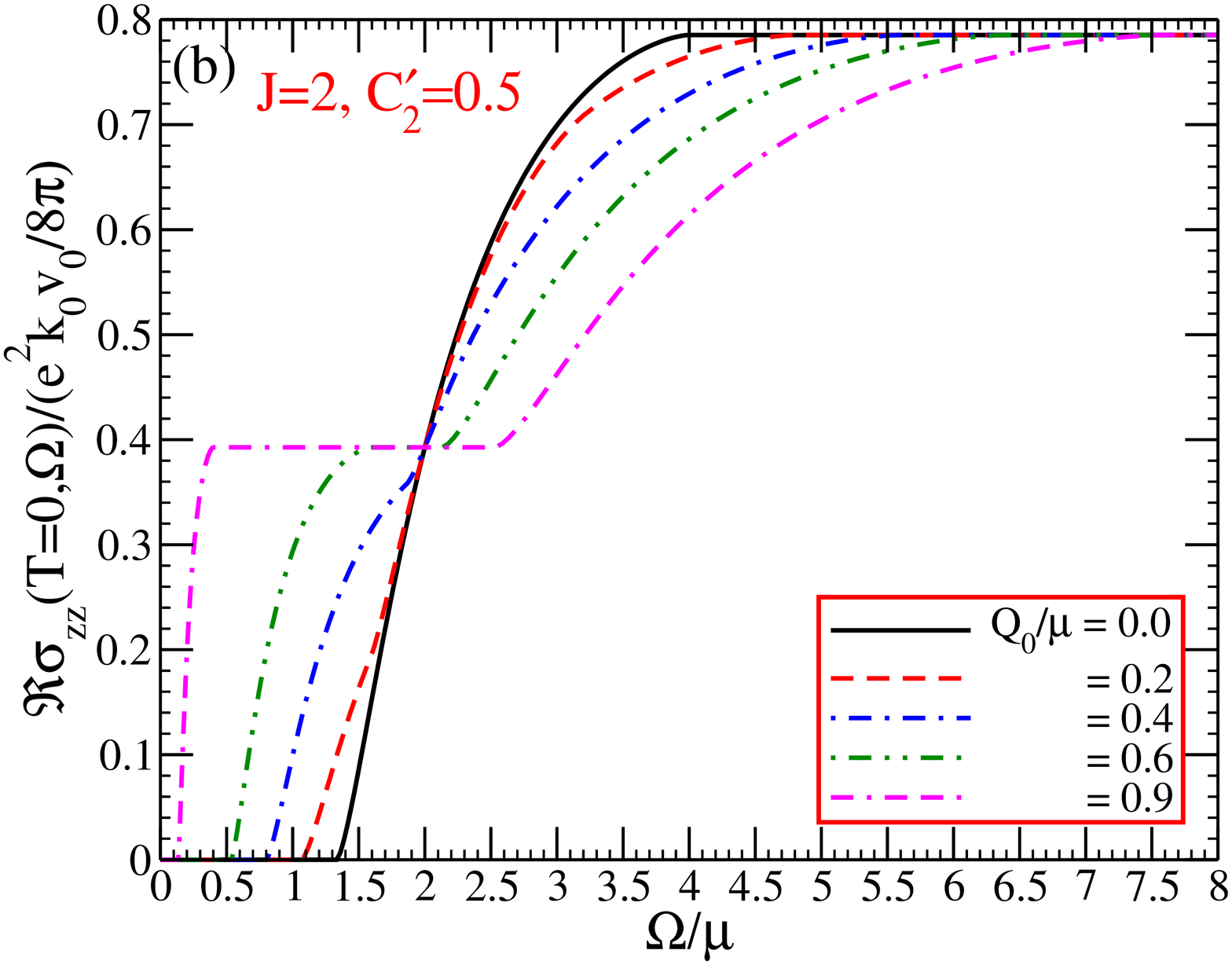}
\caption{(Color online) The real part of the dynamic optical conductivity along the $z$-axis $\Re{\sigma_{zz}}(T=0,\Omega)$ at zero temperature in units of $e^2\mu'v^2_{0}/8\pi$ (for $J=1$) and 
$e^2k_{0}v_{0}/8\pi$ (for $J=2$) as a function of photon energy (normalized to the chemical potential $\mu$) up to $\Omega/\mu=8$. The top frame (a) is for $J=1$ and the bottom frame (b) is for $J=2$. A 
specific value of tilt $C'_{2}=0.5$ (type I) is chosen and the inversion symmetry breaking parameter $Q_{0}/\mu$ is changed. Five values are chosen $Q_{0}/\mu=0$ (solid black), $Q_{0}/\mu=0.2$ (dash red),
$Q_{0}/\mu=0.4$ (dash-dot blue), $Q_{0}/\mu=0.6$ (dash double dot green) and $Q_{0}/\mu=0.9$ (dot double dash magenta).} 
\label{Fig8}
\end{figure}
In Fig.[\ref{Fig7}] we presents the same kind of results as for the lower frame of Fig.[\ref{Fig5}] and [\ref{Fig6}] but now $C'_{2}$ is fixed at a value of $C'_{2}=0.5$ and $Q_{0}/\mu$ is varied with 
$J=1$. The top frame gives results for $\Re{\sigma_{xx}}(T=0,\Omega)$ and the bottom for $\Re{\sigma_{zz}}(T=0,\Omega)$. Five values of $Q_{0}/\mu$ are considered. The solid black is for $Q_{0}/\mu=0$, 
dash red for $Q_{0}/\mu=0.2$, dash-dot blue for $Q_{0}/\mu=0.4$, dash double dot green for $Q_{0}/\mu=0.6$ and dot double dash magenta for $Q_{0}/\mu=0.9$. For $Q_{0}/\mu=0$ solid black both Weyl nodes 
contribute equally to the absorptive part of the conductivity. The main difference between top and bottom curve is the rise out of zero absorption at $\Omega/\mu=1.33$. It is sharper and concave down in 
$\Re{\sigma_{xx}}(T=0,\Omega)$ than for $\Re{\sigma_{zz}}(T=0,\Omega)$ which is concave upward, but these are small differences and both curves are qualitatively quasilinear with slightly different 
slopes. As $Q_{0}/\mu$ is increased the absorption starts at ever decreasing values of photon energies $\Omega/\mu=\frac{2(1-Q_{0}/\mu)}{1+C'_{2}}$ and a region develops which is entirely due to the negative 
chirality node up to $\frac{2(1+Q_{0}/\mu)}{1+C'_{2}}$. Such a region never arises in centrosymmetric materials. Further in the dash double dot green curve for $Q_{0}/\mu=0.6$ and the dot double dash magenta 
for $Q_{0}/\mu=0.9$ the negative chirality node contribution to the conductivity has recovered its no-tilt slope of $1/3$ as is clearly seen in the figure. This
\begin{figure}[H]
\centering
\includegraphics[width=2.5in,height=3.0in, angle=270]{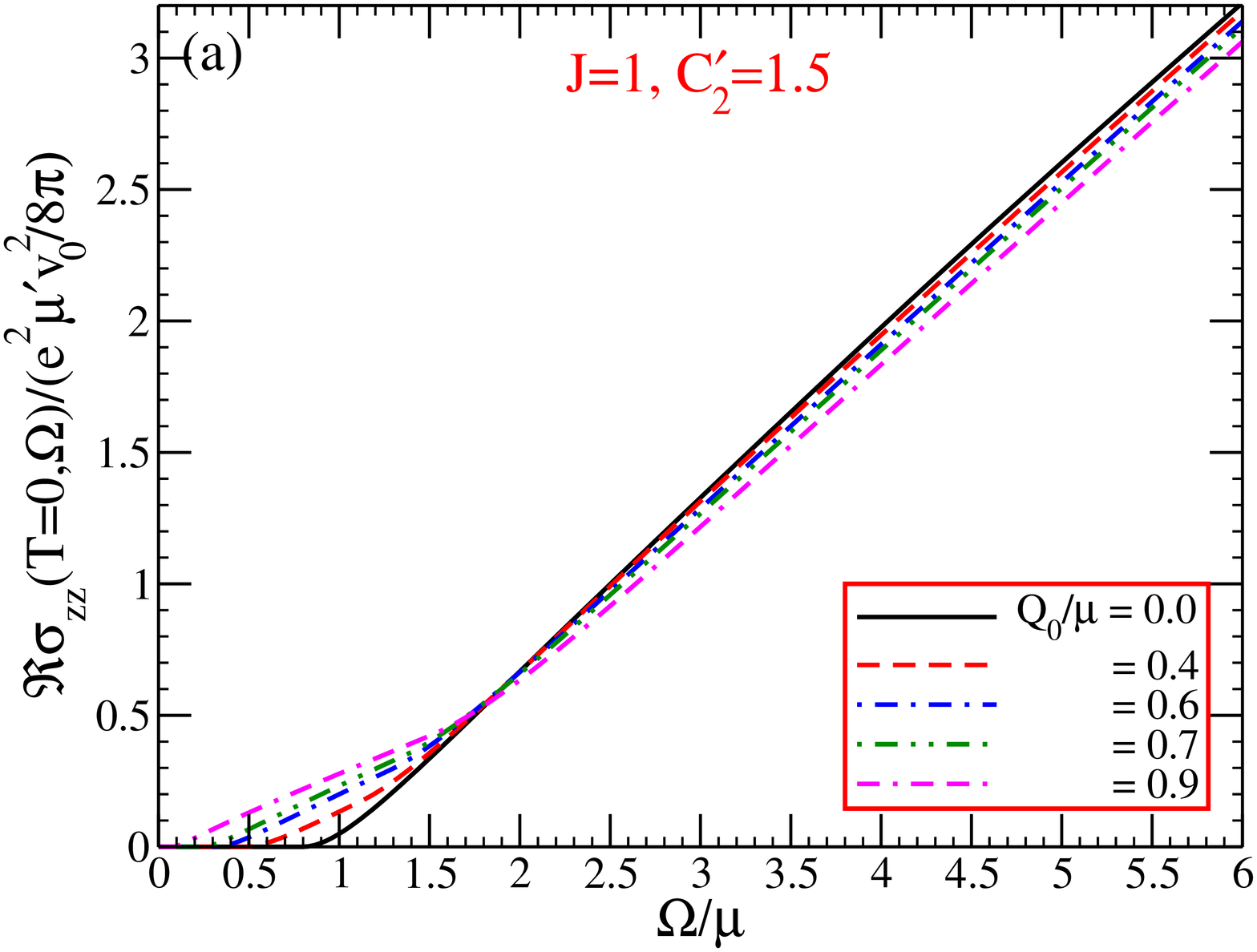}
\includegraphics[width=2.5in,height=3.0in, angle=270]{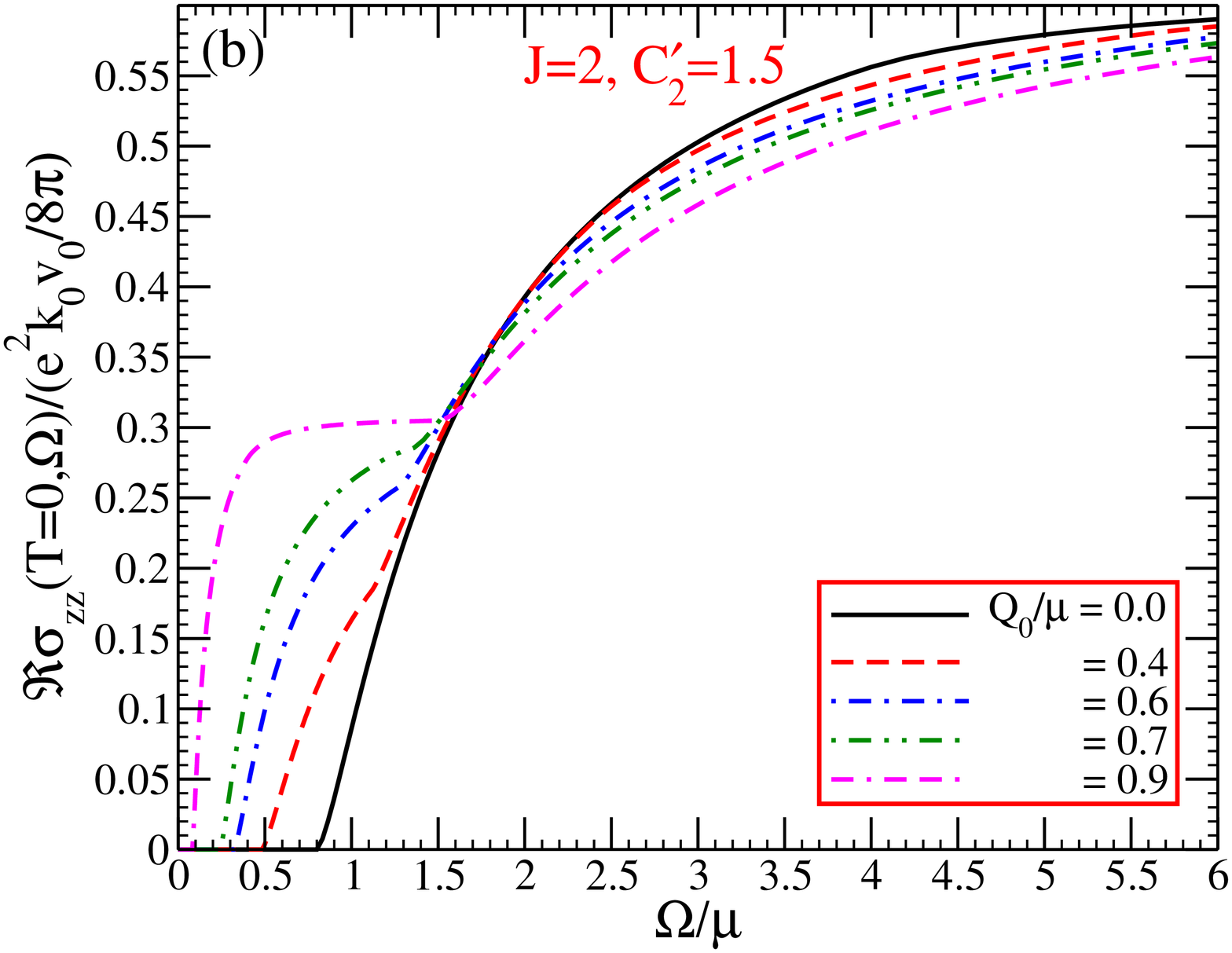}
\caption{(Color online) The real part of the dynamic optical conductivity along the $z$-axis $\Re{\sigma_{zz}}(T=0,\Omega)$ at zero temperature in units of $e^2\mu'v^2_{0}/8\pi$ (for $J=1$) and 
$e^2k_{0}v_{0}/8\pi$ (for $J=2$) as a function of photon energy (normalized to the chemical potential $\mu$) up to $\Omega/\mu=6$. The top frame (a) is for $J=1$ and the bottom frame (b) for $J=2$. A 
specific value of tilt $C'_{2}=1.5$ (type II) is chosen and the inversion symmetry breaking parameter $Q_{0}/\mu$ is changed. Five values are chosen $Q_{0}/\mu=0$ (solid black), $Q_{0}/\mu=0.4$ (dash red),
$Q_{0}/\mu=0.6$ (dash-dot blue), $Q_{0}/\mu=0.7$ (dash double dot green) and $Q_{0}/\mu=0.9$ (dot double dash magenta).} 
\label{Fig9}
\end{figure}
\noindent
perfectly linear region extends from $\Omega/\mu=1.6$ to 2.13 for $Q_{0}/\mu=0.6$ and from 0.4 to 2.53 for $Q_{0}/\mu=0.9$.

In Fig.[\ref{Fig8}] and [\ref{Fig9}] we compare results for the $\Re{\sigma_{zz}}(T=0,\Omega)$ in units of $e^2\mu'v^2_{0}/8\pi$ and $e^2 k_{0}v_{0}/8\pi$ top and bottom frames respectively, for two 
values of tilt $C'_{2}=0.5$ (type I) in Fig.[\ref{Fig8}] and 
$C'_{2}=1.5$ (type II) in Fig.[\ref{Fig9}]. The top frames are for winding number $J=1$ and the bottom for $J=2$ (multi-Weyl). The value of the inversion symmetry breaking parameter $Q_{0}$ which 
corresponds to the energy shift $\pm Q_{0}$ associated with the two Weyl nodes is varied. In Fig.[\ref{Fig8}] $Q_{0}/\mu=0$ (solid black), $Q_{0}/\mu=0.2$ (dash red), $Q_{0}/\mu=0.4$ (dash-dot blue), 
$Q_{0}/\mu=0.6$ (dash double dot green) and $Q_{0}/\mu=0.9$ (dot double dash magenta). In Fig.[\ref{Fig9}] the line type and color are the same but $Q_{0}/\mu=0,0.4,0.6,0.7$ and $0.9$. In the top frame 
of Fig.[\ref{Fig8}] the slope of the high energy linear in $\Omega/\mu$ law remains $2/3$ and all the curves have merged in this region. In the solid black curve for $Q_{0}/\mu=0$ the effective 
chemical potential is the same for both nodes and they contribute equally. For $Q_{0}/\mu=0.9$ dot double dash magenta we see that a second region of linear dependence has clearly emerged and the slope 
of this line is $1/3$ because only the negative chirality node is contributing and except for very small values of $\Omega$ we are in the region $\Omega>2\mu_{-}/(1-C'_{2})$ where the no tilt law applies. 
This is a clear signature of broken inversion symmetry. The growth of $Q_{0}$ has provided the mechanism whereby the negative chirality node is separately revealed as is its characteristic slope. A 
similar situation holds for the case $J=2$ shown in the bottom frame of Fig.[\ref{Fig8}]. In the black curve both chirality nodes contribute equally. In the dot double dash magenta curve for $Q_{0}/\mu=0.9$ 
the two plateaus of height $\pi/8$ and $\pi/4$ respectively are clearly seen. For type II Weyl Fig.[\ref{Fig9}] applies and in that instance the tilt provides changes even for $\Omega/\mu=6$ where the 
slope of the high energy quasilinear behavior in the top frame is now less than $2/3$ and the various curves do not quit merge. In the lower frame two plateaus are seen but both are reduced from their 
magnitude in Fig.[\ref{Fig8}].

\section{Summary and conclusion}

Within the Kubo formulation for the AC optical conductivity of a multi-Weyl semimetal we derive analytic algebraic equations for its absorptive part both perpendicular $\Re{\sigma_{xx}}(\Omega)$ and 
parallel $\Re{\sigma_{zz}}(\Omega)$ to the symmetry axis as a function of photon energy $\Omega$. Our starting Hamiltonian contains a multi-Weyl relativistic term with winding number $J$ as well as a 
term that explicitly breaks time reversal and another inversion symmetry. An arbitrary tilt of the Weyl cones covering both type I and II is included as well as finite doping away from charge neutrality. 
Breaking time reversal symmetry splits a doubly degenerate Dirac point into two Weyl nodes of opposite chirality which are displaced in momentum space by $\pm\bf{Q}$. Breaking inversion symmetry further 
splits the nodes in energy by $\pm Q_{0}$. In the clean limit employed in this work the momentum space separation of the two Weyl nodes drops out of the absorptive part of the AC conductivity. This is in 
sharp contrast to the critical role $\bf{Q}$ plays in the anomalous DC Hall conductivity which is known\cite{Burkov,Burkov2,Tiwari} to be directly proportional to $\bf{Q}$. Terms proportional to $\bf{Q}$ 
also enter other transport coefficients\cite{Tiwari,Ferreiros,Saha}. In the case of $Q_{0}=0$ (centrosymmetric) our expressions properly reduce to known results. For winding number $J=1$ results for 
$\Re{\sigma_{xx}}(\Omega)$ when both the tilt and the doping are non-zero are given in Ref.[\onlinecite{Mukherjee1}]. Ref.[\onlinecite{Mele}] gives results in the multi-Weyl case for both 
$\Re{\sigma_{xx}}(\Omega)$ and $\Re{\sigma_{zz}}(\Omega)$ but mainly when doping and tilt are zero. Assuming the magnitude of the tilt is the same for both nodes and the system has inversion symmetry 
($Q_{0}=0$) each Weyl point contributes equally to the conductivity. For finite chemical potential $\mu$, but zero tilt there is no absorption below $\Omega=2\mu$ due to the Pauli blocking of the 
interband optical transitions. The lost optical spectral weight is transferred to the intraband Drude which, in the clean limit, is a Dirac delta function at $\Omega=0$. Above the sharp discontinuous 
absorption threshold at $2\mu$ the conductivity takes on its zero doping value. For $\Re{\sigma_{xx}}(\Omega)$ this is a linear law for both $J=1$ and $J=2$. For $\Re{\sigma_{zz}}(\Omega)$ it is again 
linear in $\Omega$ for $J=1$ but for $J=2$ it is instead constant \cite{Mele} independent of $\Omega$. A finite tilt introduces important modifications and these are qualitatively different for type I 
and type II. The boundary between these two types corresponds to the case where the Dirac cone is completely tipped over and electron-hole pockets begin to form at charge neutrality. Tilting transfers 
optical spectral weight from above the $\Omega=2\mu$ threshold to below and the sharp absorption edge of the no tilt case is lost. For type I Weyl the changes are confined to the photon range 
$2\mu/(1+C'_{2})$ to $2\mu/(1-C'_{2})$. In this range the absorption is roughly quasilinear and its value at $\Omega=2\mu$ is exactly half of its no tilt magnitude. Above $2\mu/(1-C'_{2})$, $\Re{\sigma_{xx}}(\Omega)$ 
is linear in $\Omega$ for both winding number $J=1$ and 2 as is $\Re{\sigma_{zz}}(\Omega)$ for $J=1$. For $J=2$, $\Re{\sigma_{zz}}(\Omega)$ is instead constant. For type II Weyl, the changes due to 
tilting are more pronounced and they extend to large values of $\Omega$ where the absorption never returns to its zero tilt value.

For noncentrosymmetric multi-Weyl semimetals the nodes are no longer at the same energy and effectively each Weyl cone has a different value of chemical potential ($\mu_{s'}=\mu+s'Q_{0}$) with $s'=\pm$ 
(positive/negative) chirality. For the no tilt case there is now two sharp absorption edges and a region of photon energy emerges between the two absorption thresholds for which only the negative chirality 
node contributes. The magnitude of the energy shift $Q_{0}$ controls the size of this region in photon energy. For finite tilt both absorption edges are modified as we previously described for the $Q_{0}=0$ 
case. The positive chirality node contributes only for photon energy greater than $2\mu(1+Q_{0}/\mu)/(1+C'_{2})$. For type I Weyl this can be larger than the value of $\Omega$ at which the negative 
chirality node has recovered its no tilt behavior. This occurs from $Q_{0}/\mu>C'_{2}$. In this situation there is an interval of photon energy where not only is the absorption due only to the negative 
chirality node but it also takes on its no tilt behavior. For $\Re{\sigma_{xx}}(\Omega)$ this is $\frac{e^2}{8\pi}\frac{\Omega}{3v_{z}}J$, for $\Re{\sigma_{zz}}(\Omega)$ with $J=1$ it is 
$\frac{e^2}{8\pi}\frac{\Omega}{3v_{z}}v^2_{0}$ while with $J=2$ it is $\frac{e^2}{8\pi}k_{0}v_{0}$. Above $2\mu(1+Q_{0}/\mu)/(1-C'_{2})$ for type I, both Weyl nodes contribute equally to the conductivity 
and its value is exactly twice the  value quoted above. For type II in the limit of very large tilt and $\Omega$ also large, the previous laws still hold but the numerical factors are changed and there 
is an extra factor of $1/C'_{2}$ which reduces the slopes to zero for the linear laws and the magnitude of the constant background for $\Re{\sigma_{zz}}(\Omega)$ with $J=2$.

\subsection*{Acknowledgments}
Work supported in part by the Natural Sciences and Engineering Research Council of Canada (NSERC)(Canada) and by the Canadian Institute for Advanced Research (CIFAR)(Canada). We thanks A.A.Burkov and 
D. Xiao for enlightening discussions.

\begin{widetext}
\appendix

\section{}

Following Eq.(\ref{sigma-xx}) the real part of $\sigma_{xx}(\Omega)$ is written as,
\bea
&& \Re\sigma_{xx}(\Omega)=-\frac{e^2J^2 k^{-(2J-1)}_{0}v^2_{\perp}}{2\pi\Omega}\sum_{s'=\pm}\int^{\Lambda-s'Q}_{-\Lambda-s'Q} dk_{z}\int^{\infty}_{0} \frac{k^{2J-1}_{\perp}dk_{\perp}}{k}\{f(C_{s'}k_{z}+k_{0}k-\mu_{s'})-f(C_{s'}k_{z}-k_{0}k-\mu_{s'})\} \times \nonumber\\
&& (k^2_{0}k^2+v^2_{z}k^2_{z}) \delta(4k^2_{0}k^2-\Omega^2) \nonumber\\
&&=-\frac{e^2J^2 k^{-2J}_{0}v^2_{\perp}}{8\pi\Omega^2}\sum_{s'=\pm}\int^{\Lambda-s'Q}_{-\Lambda-s'Q} dk_{z}\int^{\infty}_{0} \frac{k^{2J-1}_{\perp}dk_{\perp}}{k}\{f(C_{s'}k_{z}+k_{0}k-\mu_{s'})-f(C_{s'}k_{z}-k_{0}k-\mu_{s'})\} \times \nonumber\\
&& (k^2_{0}k^2+v^2_{z}k^2_{z}) \lb \delta(k-\frac{\Omega}{2k_{0}})+\delta(k+\frac{\Omega}{2k_{0}})\rb\nonumber\\
&& =-\frac{e^2J}{8\pi\Omega^2}\sum_{s'=\pm}\int^{\Lambda-s'Q}_{-\Lambda-s'Q} dk_{z}\int^{\infty}_{v_{z}|\frac{k_{z}}{k_{0}}|} dk \{f(C_{s'}k_{z}+k_{0}k-\mu_{s'})-f(C_{s'}k_{z}-k_{0}k-\mu_{s'})\} (k^2_{0}k^2+v^2_{z}k^2_{z})\delta(k-\frac{\Omega}{2k_{0}})\nonumber\\
&& \text{Taking the cut off $\Lambda$ to be much larger than the momentum space separation between the Weyl nodes and also } \nonumber \\
&& \text{much larger than $\Omega/2k_{0}$ we get,} \nonumber\\
&& = -\frac{e^2Jk_{0}}{32\pi v_{z}\Omega^2}\sum_{s'=\pm}\int^{\frac{\Omega}{2k_{0}}}_{-\frac{\Omega}{2k_{0}}} dk_{z} \lf f(\frac{C_{s'}k_{0}}{v_{z}}k_{z}+\frac{\Omega}{2}-\mu_{s'})-f(\frac{C_{s'}k_{0}}{v_{z}}k_{z}-\frac{\Omega}{2}-\mu_{s'})\rf (\Omega^2+4k^2_{0}k^2_{z})
\eea
We note that the variable $Q$ has entirely dropped out of this quantity. Here we can drop the second Dirac delta function as it clicks at $k=-\frac{\Omega}{2k_{0}}$ which is outside the range of integration. 
In the second step we have changed the variable $k_{\perp}$ to $k$ as shown below,
\bea
&& k^2= v^2_{\perp}(k_{\perp}/k_{0})^{2J}+v^2_{z}(k_{z}/k_{0})^2 \nonumber\\
&& kdk=\frac{Jv^2_{\perp}}{k_{0}}(k_{\perp}/k_{0})^{2J-1} dk_{\perp}\nonumber\\
\eea
In the limit $T\to 0$ the Fermi functions become theta functions. This gives,
\bea
&& \hspace{-1.0cm}\Re\sigma_{xx}(\Omega)=-\frac{e^2J v^2_{z}}{8\pi \Omega^2}\sum_{s'=\pm}\hspace{-0.1cm}\int^{\frac{\Omega}{2v_{z}}}_{0} \hspace{-0.2cm} dk_{z} \biggl[ \Theta(C_{s'}k_{z}-\frac{\Omega}{2}-\mu_{s'}) - 
\Theta(C_{s'}k_{z}+\frac{\Omega}{2}-\mu_{s'})- \Theta(-C_{s'}k_{z}+\frac{\Omega}{2}-\mu_{s'})\biggr] (k^2_{z}+\frac{\Omega^2}{4v^2_{z}})
\eea
We see that the above expression for $\Re\sigma_{xx}(\Omega)$ is tilt-inversion symmetric i.e. if we change $C_{s'}$ to $-C_{s'}$ then $\Re\sigma_{xx}(\Omega)$ stays same. It only depends on the absolute 
value of the tilts in two Weyl nodes irrespective of whether the tilt is clockwise or anticlockwise. The rest of the calculation is straight forward and we state the final result in the main text in Sec.\ref{sec:III}.

\section{}

Here we derive the result for the conductivity $\sigma_{zz}(\Omega)$. Following Eq.(\ref{sigma-zz})the real part of $\sigma_{zz}(\Omega)$ is written as,
\bea
&& \Re\sigma_{zz}(\Omega)=-\frac{e^2}{8\pi k^2_{0}\Omega^2}\sum_{s'=\pm}\int^{\Lambda-s'Q}_{-\Lambda-s'Q} dk_{z}\int^{\infty}_{0} \frac{k_{\perp} dk_{\perp}}{k}\{f(C_{s'}k_{z}+k_{0}k-\mu_{s'})-f(C_{s'}k_{z}-k_{0}k-\mu_{s'})\} \times \nonumber \\
&& \lb (C^2_{s'}+v^2_{z})\lf k^2_{0}k^2-v^2_{z}k^2_{z}\rf-v^2_{\perp}(C^2_{s'}-v^2_{z})k^{2J}_{\perp}k^{-2(J-1)}_{0}\rb \lb \delta(k-\frac{\Omega}{2k_{0}})+\delta(k+\frac{\Omega}{2k_{0}})\rb\nonumber\\
&& =-\frac{e^2k^2_{0}v^2_{z}}{4\pi J(k^2_{0}v^2_{\perp})^\frac{1}{J}\Omega^2}\sum_{s'=\pm}\int^{\Lambda-s'Q}_{-\Lambda-s'Q} dk_{z}\int^{\infty}_{v_{z}|\frac{k_{z}}{k_{0}}|} dk \lf f(C_{s'}k_{z}+k_{0}k-\mu_{s'})-f(C_{s'}k_{z}-k_{0}k-\mu_{s'})\rf \lp k^2_{0}k^2-v^2_{z}k^2_{z}\rp^\frac{1}{J} \times \nonumber\\
&& \delta(k-\frac{\Omega}{2k_{0}})\nonumber\\
&& =-\frac{e^2k^2_{0}v^2_{z}}{4^{\frac{1+J}{J}}\pi J(k^2_{0}v^2_{\perp})^\frac{1}{J}\Omega^2}\sum_{s'=\pm}\int^{\frac{\Omega}{2v_{z}}}_{-\frac{\Omega}{2v_{z}}} dk_{z}\lf f(C_{s'}k_{z}+\frac{\Omega}{2}-\mu_{s'})-f(C_{s'}k_{z}-\frac{\Omega}{2}-\mu_{s'})\rf \lp \Omega^2-4v^2_{z}k^2_{z}\rp^\frac{1}{J}
\eea
At zero temperature limit we get,
\bea
&& \hspace{-0.4cm}\Re\sigma_{zz}(\Omega)=-\frac{e^2k^2_{0}v^2_{z}}{4^{\frac{1+J}{J}}\pi J(k^2_{0}v^2_{\perp})^\frac{1}{J}\Omega^2}\sum_{s'=\pm}\int^{\frac{\Omega}{2v_{z}}}_{0} dk_{z} \biggl[ \Theta(C_{s'}k_{z}-\frac{\Omega}{2}-\mu_{s'}) - \Theta(C_{s'}k_{z}+\frac{\Omega}{2}-\mu_{s'})-\Theta(-C_{s'}k_{z}+\frac{\Omega}{2}-\mu_{s'})\biggr] \nonumber \\
&& \times \lp \Omega^2-4v^2_{z}k^2_{z}\rp^\frac{1}{J}
\eea

\end{widetext}

\end{document}